\documentclass[a4paper,5p,times,sort&compress]{elsarticle}
\usepackage[utf8]{inputenc}
\usepackage[T1]{fontenc}

\graphicspath{{graphics/}}

\usepackage{amsmath}
\usepackage{amsfonts}
\usepackage{amssymb}
\usepackage{float}
\usepackage[normalem]{ulem}

\usepackage{booktabs}
\usepackage{tabularx}
\usepackage{threeparttable}
\usepackage{rotating}
\usepackage{siunitx}

\usepackage{verbatim}

\usepackage{mwe}
\usepackage{multirow}

\PassOptionsToPackage{hyphens}{url}\usepackage[colorlinks=true, citecolor=blue, linkcolor=blue, filecolor=blue,urlcolor=blue]{hyperref}

\usepackage[gen]{eurosym}

\usepackage{lineno}
\modulolinenumbers[5]

\usepackage{framed} %

\usepackage{multicol} %

\usepackage{nomencl} %

\usepackage{xcolor}

\makenomenclature

\renewcommand*\nompreamble{\begin{multicols}{2}}

\renewcommand*\nompostamble{\end{multicols}}

\journal{}

\begin{document}

\begin{frontmatter}

\title{Open-data based carbon emission intensity signals for electricity generation in European countries -- top down vs. bottom up approach}

\author[INATECH]{Jan~Frederick~Unnewehr\corref{cor1}}
\ead{jan.frederick.unnewehr@inatech.uni-freiburg.de}

\author[INATECH]{Anke~Weidlich}
\author[INATECH]{Leonhard~Gf\"{u}llner}
\author[INATECH]{Mirko~Sch\"{a}fer}

\cortext[cor1]{Corresponding author}

\address[INATECH]{INATECH, University of Freiburg, 79110, Germany}

\begin{abstract}
Dynamic grid emission factors provide a temporally resolved signal about the carbon intensity of electricity generation in the power system. Since actual carbon dioxide emission measurements are usually lacking, such a signal must be derived from  system-specific emission factors combined with power generation time series. We present a bottom-up method that allows deriving per country and per technology emission factors for European countries based on plant specific power generation time series and reported emissions from the European emissions trading mechanism. We have matched, 595~fossil generation units and their respective annual emissions. In 2018, these power plants supplied 717\,TWh of electricity to the grid, representing approximately 50\,\% of power generation from fossil fuels. Based on this dataset, 42 individual technology and country-specific emission factors are derived. The resulting values for historical per country carbon intensity of electricity generation are compared with corresponding results from a top-down approach, which uses statistical data on emissions and power generation on national scales. All calculations are based on publicly available data, such that the analysis is transparent and the method can be replicated, adjusted and expanded in a flexible way.
\end{abstract}


\end{frontmatter}


\section{Introduction}
\label{sec:intro}

Decarbonizing the power sector plays a key role in Europe’s ambition to be the first climate-neutral continent by 2050~\cite{EU2019, EU2020}. For the required reduction of carbon dioxide emissions in the electricity system, it is crucial to make detailed and transparent emission data available to consumers, regulators, and other stakeholders~\cite{Hamels2021}. Regulators, for instance, must track the average emission intensity of the electricity mix in their domain to monitor the fulfillment of given climate targets. Meanwhile, dynamic grid electricity emission measures with a high temporal resolution provide a signal for smart energy systems with storage or demand side flexibility. Such systems, ranging from data centers or urban districts to charging networks for electric vehicles, rely on dynamic grid electricity emission signals to optimize their operational schedule aiming for a minimum carbon footprint~\cite{Clauss2019,Alavijeh2020,Google2021,RMI2021}.

Since actual comprehensive emission measurements are not yet available~\cite{climatetrace}, the development of emissions in the power sector is typically tracked through the calculation of total emissions based on emission factors (EFs). The EF is the quotient that relates the amount of a pollutant (e.g., carbon dioxide emissions) released into the atmosphere to an activity (e.g., production of one MWh of electricity) associated with the release of that pollutant~\cite{EEA2021}. 

One of the most prominent applications is the emission determination for reporting emissions under the United Nations Framework Convention on Climate Change (UNFCCC) \cite{1992UNITEDNATIONS}. Under this convention, all participating countries must prepare and submit a National Inventory Report for their greenhouse gas emissions based on one of three methodological approaches.

Tier~1 uses default EFs provided in the IPCC guidelines. The method can be combined with spatially explicit activity data from remote sensing, for example satellite data. Tier~2 generally uses the same methodological approach as tier~1, but applies EFs that are country specific. Country-specific EFs are those that are more appropriate for the climate regions and systems used in that country. At tier~3, more complex methods, including measurements and analytical models to address national circumstances are used~\cite{Eggleston20062006Inventories}.

In life-cycle assessments (LCA), similar  approaches are used for assessing the carbon intensity of energy and material inputs to a product, process or service. In LCA, it is common to use an annual mean value that refers to the average emissions that are caused by the production of one MWh of final energy~\cite{Turconi2013LifeLimitations}. The corresponding calculations are usually based on the energy conversion processes of representative power plants, from which emissions for specific technologies in one or more countries are derived. Generally, this approach does not adequately represent the variability of EFs in the generation fleet. As a result, it is difficult to compare the resulting emissions of different countries \cite{Hamels2021, Turconi2013LifeLimitations}. In many studies, this emission value is often denoted as the grid-based carbon intensity (CI) of a given country. In most cases of grid-based CIs, annual emissions are divided by the annual electricity production, which leads to a static average emission value per year. This approach to evaluate emissions from different sectors is similar to the tier~2 approach from the IPCC guideline~\cite{Astudillo2017LifeOpportunities,Itten2012LifeGrid}.

Several studies have shown that there are significant variations in the electricity mix from month to month, and even hour to hour variations within a day~\cite{Tranberg2019Real-timeMarkets}. These variations are caused by fluctuations in the contribution of both conventional and renewable power generation. Consequently, the emissions associated with the use of electricity also vary across time~\cite{Spork2015IncreasingFactors,Noussan2018PerformanceItaly,Kopsakangas-Savolainen2017Hourly-basedEmissions,Vuarnoz2018TemporalGrid,Marrasso2019ElectricBasis}. This shows that reporting schemes based on aggregated emissions and power generation are not suitable to provide emission intensity signals on hourly time scales. However, this timely resolved information is crucial for the operation of flexible systems, like charging stations for electric vehicles, heat pumps to data centers, are to be optimized with respect their associated carbon dioxide emissions~\cite{RMI2021}. 

The provision of a widely applicable and transparent carbon intensity signal is facilitated by the incorporation of publicly available consistent and consolidated data, as well as the use of a transparent and easy-to-use method for determining CO$_2$ intensities~\cite{Hamels2021}. Currently, the regulations and scientific research in the area of energy system analysis often use standard EFs that are neither technology nor country specific. These standard EFs do not fulfill either of the criteria 'applicability', 'accuracy' and 'transparency' sufficiently. Often, the data sources used are not openly available, and the underlying calculations are hard to replicate due to insufficient documentation. Additionally, the methods used vary from country to country, making it difficult to compare the resulting EFs or CIs. 

To address this gap, in this study we introduce a consistent and flexible framework to calculate EFs based on per unit generation and emission data. Through this bottom-up approach, emissions can be assessed from the unit up to country level on different time scales, depending on the spatial and temporal resolution of the given generation data. The derived values for the average carbon intensity of electricity generation for European countries are compared to results calculated through a top-down approach based on national statistics~\cite{EEA2021}. The accuracy of the methods and the need of further data consolidation is discussed. The key novelty of this study is the development and assessment of such a self-contained, modular bottom-up approach for calculating EFs based on publicly available data only. Through the provisioning of all code and secondary data, the calculation framework is transparent and accessible for modifications and extensions, providing a solid foundation for future studies.

The remainder of this article is structured as follows. Section~\ref{sec:back_and_lit} provides an overview of different calculation schemes and applications of EFs. In Section~\ref{sec:data}, the data sources underlying this study are reviewed, followed by the presentation of the bottom-up and top-down method in Section~\ref{sec:method}. Results are presented in Section~\ref{sec:application}, with a subsequent discussion in Section~\ref{sec:discussion}. Section~\ref{sec:conclusion} concludes this article.

\section{Research background}
\label{sec:back_and_lit}

A brief literature overview of the calculation and application of EFs is given in the following. This is complemented by a summary of the different methods used to model EFs and resulting CIs, including use cases in the literature in Table~\ref{tab:literature_review}. 

\begin{table*}[]
\begin{tabular}{@{}p{3.3cm}p{3.6cm}p{3cm}llp{3.5cm}@{}}
\toprule
Source & EF calculation method & EF properties & \multicolumn{2}{c}{CI resolution} & Application \\ 
 &   &  & Temporal & Spatial & \\ \midrule
\multicolumn{2}{l}{Calculation-based EF}   &   &   &  &  \\ \midrule Braeuer et al. \cite{Braeuer2020ComparingGermany} & Empirical emission \& generation data & Tech. specific, direct emissions & hourly & Germany & Battery storage  dispatch optimization \\
Staffell et al. \cite{Staffell2017MeasuringElectricity} & Carbon content of fuel & Tech. specific, direct emissions & yearly & Great Britain & Decarbonisation progress measurement \\
Hein et al. \cite{Hein2020Agorameter-Dokumentation} & Carbon content of fuel & Tech. specific, direct emissions & yearly & Germany & Emission visualisation \\ \midrule
\multicolumn{2}{l}{Literature-based EF}   &   &   &  &  \\ \midrule
Spork et al. \cite{Spork2015IncreasingFactors} & Literature-based (IEA) & Tech. specific, direct emissions & hourly & Spain & Company dispatch optimization \\
Marrasso et al. \cite{Marrasso2019ElectricBasis} & Literature-based (ISPRA) & Tech. specific, direct emissions & hourly & Italy & Decarbonisation progress measurement \\
Noussan et al. \cite{Noussan2018PerformanceItaly} & Literature-based (ISPRA) & Country specific, direct emissions & hourly & Italy & Decarbonisation progress measurement \\
Dixit et al. \cite{Dixit2014CalculatingSectors} & Literature-based & Country specific, direct emissions & yearly & United States & Decarbonisation progress measurement \\
Vuarnoz et al. \cite{Vuarnoz2018TemporalGrid} & Literature-based (LCA) & Tech. specific, direct and indirect emissions & hourly & Switzerland & LCA application \\
Moro and Lonza \cite{Moro2018ElectricityVehicles} & Literature-based (LCA) & Tech. specific, direct and indirect emissions & yearly & Europe & Electric vehicle charging \\
Kopsakangas-Savolaine et. al. \cite{Kopsakangas-Savolainen2017Hourly-basedEmissions} & Literature-based & Tech. specific, direct emissions & hourly & Finland & Household consumption optimization \\
Arciniegas and Hittinger \cite{Arciniegas2018TradeoffsOperation} & Literature-based & Country specific, direct emissions & hourly & United States & Battery storage dispatch optimization \\
Ang and Su \cite{Ang2016CarbonAnalysis} & Literature-based & Country specific, direct emissions & yearly & World & Decarbonisation progress measurement \\
Tranberg et. al. \cite{Tranberg2019Real-timeMarkets} & Literature-based (LCA) & Tech. specific, direct and indirect emissions & yearly & Europe & Electricity market carbon accounting  \\
Qu et. al. \cite{Qu2017} & Literature-based & Country specific, direct emissions & yearly & Eurasia & Electricity trade carbon accounting \\
Zafiraki et. al. \cite{Zafirakis2015} & Literature-based (IEA) & Tech. specific, direct emissions & monthly & Europe & Storage dispatch optimization \\
\bottomrule
\end{tabular}
\caption{Summary of the methods used for calculating emission factor and the use cases presented in the literature.}
\label{tab:literature_review}
\end{table*}

\subsection{Calculation of emission factors}

Most approaches to calculating EFs follow either a process-based (LCA) scheme, or a balancing method in which total emissions and total generated electricity are used.

The process-based LCA methods derive the resulting emissions from a systematic analysis of the underlying processes and infrastructure. In LCA studies, total emissions are usually classified as either direct emissions (e.g., from the combustion of the fuel) or indirect emissions (e.g., related to upstream fuel supply, resources, construction of the power plant, etc.)~\cite{Turconi2013LifeLimitations}. Many LCA studies can be found for different power generation technologies. In~\cite{Odeh2008LifePlants}, for instance, the authors derive an EF of 882\,gCO$_2$/kWh for coal-fired power plants associated with direct emissions resulting from fuel combustion. 

In~\cite{Turconi2013LifeLimitations}, the authors examine 167 LCA studies of electricity generation for various technologies. One key finding of the analysis is the broad range of literature values for EFs for different generation technologies, showing a strong dependence on parameters such as location, efficiency, fuel quality, or type of use.  With respect to direct emissions, EFs range from 660 to 1\,050\,gCO$_2$/kWh for hard coal-fired power plants, whereas for lignite power plants, values between 800 and 1\,300\,gCO$_2$/kWh can be found. For gas power plants, EFs vary between 380 and 1\,000\,gCO$_2$/kWh. 

The balancing approach determines EFs via substance and energy flows. EFs for energy systems are often based on data for the power generation as well as the amount of fuel used to generate the respective power. Furthermore, a fuel-specific EF is required. These factors can be obtained from chemical fuel analyses. In this method, the fuel quantity is multiplied by the fuel-specific EF, and divided by the respective energy yield. The resulting value represents the technology-specific EF. For instance, in~\cite{Hein2020Agorameter-Dokumentation}, data from the Federal Environment Agency of Germany~\cite{Umweltbundesamt2021Entwicklung2020} is used to determine a technology-specific EF via substance flows. These factors are combined with hourly generation data to provide a CO$_2$ signal of the German electricity mix with hourly resolution. Nevertheless, since the input data is unavailable in a technology-specific manner, only country-specific values are determined.

In~\cite{Braeuer2020ComparingGermany}, two approaches for calculating a dynamic grid EF are compared. Different to other studies, which in general include reported annual power generation and emissions, the authors pair plant emission data from the European Union Emissions Trading System (EU~ETS) with generation per unit data from \mbox{ENTSO-E} to calculate power plant-specific EFs. However, the study is limited to Germany, and the resulting EFs are not evaluated for consistency and verification based on further data sources.

\subsection{Application of emission factors}

Technology-specific EFs are widely used in the literature for various applications. In~\cite{Marrasso2019ElectricBasis}, Marrasso et al. present performance indicators for country-wide power systems that are based on annual technology-specific EFs for the Italian power system. In contrast, Noussan et al.~\cite{Noussan2018PerformanceItaly} calculate similar performance indicators for the Italian power system, but use yearly aggregated EFs. In both studies, EFs are taken from the literature, and hourly CO$_2$ emission signals are derived.

In~\cite{Vuarnoz2018TemporalGrid}, EFs per technology are used to calculate hourly greenhouse gas emissions of the national electricity supply mix in Switzerland and neighboring countries. The applied per technology EFs are based on the life-cycle inventory data of the ecoinvent~2.2 database and include transport as well as distribution losses.
In~\cite{Dixit2014CalculatingSectors}, the authors review current methods for calculating the primary energy use and the carbon emissions associated with electricity consumption. For calculating the carbon emissions, they use different technology-specific EFs provided by the U.S. Energy Information Administration~\cite{USEIA}. In~\cite{Zafirakis2015}, the authors use technology-specific EFs to determine monthly grid EFs for European countries. They investigate how the use of storage can help to reduce CO$_2$-intensive electricity trade from other countries.

In~\cite{Qu2017}, the authors introduce an easy to adapt model, that uses economic input-output theory to access carbon intensities of electricity consumption taking trade between countries into consideration. For a case study covering the Eurasia area, they used country-specific EFs and generation data to show that excluding electricity exchanges with other countries can have a significant impact on emissions estimates.

The “electricityMap” project displays hourly EFs associated with electricity generation for various countries~\cite{electricitymap}. The calculation for the European region is based on generation per production type data from various sources, predominantly \mbox{ENTSO-E}. These generation time series are multiplied by carbon emission intensities mostly taken from the IPCC (2014) Fifth Assessment Report~\cite{electricitymapgithub,IPCC2014}. Also, a consumption-based carbon intensity of electricity is derived, which uses a tracing approach to consider imports through the power grid~\cite{Tranberg2019Real-timeMarkets,deChalendar2019,Schaefer2020}.

\section{Used data and preparation}
\label{sec:data}

Generally, data on power generation has to be associated with data representing the corresponding emissions to calculate the carbon intensity and EFs. For the different methods presented here, generation and emission datasets on different scales from multiple sources are used. For the generation side, these are production time series for individual generation units (Sec.~\ref{sec:entsoe_unit}), for each production type and country (cp. Section~\ref{sec:entsoe_prod}) and consolidated aggregated yearly production per type and country (cp. Section~\ref{sec:entsoe_sheet}), all from \mbox{ENTSO-E}, as well as reported energy balances per country from Eurostat (cp. Section~\ref{sec:balance}). The included emission datasets comprise yearly emissions for individual power plants from the EU~ETS scheme (Sec.~\ref{sec:euets}) and reported annual emissions for individual countries  as part of the UNFCCC framework (cp. Section~\ref{sec:unfccc}). Table~\ref{tab:used_data} gives an overview of the used datasets and their properties.

It should be emphasized that both generation and emission datasets are inconsistent across different sources and across different scales. For instance, aggregated per unit generation time series do not match with per production type time series; aggregated production type time series from \mbox{ENTSO-E} do no match with  the reported consolidated annual values in the \mbox{ENTSO-E} fact sheet, which in turn does not accord with the energy balances published by Eurostat. Analogously, annually reported emissions per sector and country differ from aggregated reported emissions per power plants. All these inconsistencies originate from different reporting schemes, coverage and definitions, misallocation or gaps and errors in the data. Providing consistent open datasets for generation and emissions on different temporal and spatial scales, based on the sources reviewed below, is a much-needed endeavor, but beyond the scope of this study. Consequently, in our approach we aim to incorporate different datasets in separate parts of the method to reduce uncertainties originating from the inconsistencies outlined above. In the following, the datasets applied here, and the data processing are described in further detail.
\begin{table*}[]
\begin{tabular}{@{}p{3.5cm}llp{1.3cm}p{1.9cm}p{4.3cm}l}
\toprule
Name & Type & Unit & \multicolumn{2}{c}{Resolution} & Description & Reference \\
& & & Temporal & Spatial & & \\\midrule
\mbox{ENTSO-E} generation per block-unit & Generation & MW & 15\,min to hourly & Power plant unit & Net electricity production time series for large power plants 
& \cite{2020ENTSO-EUnit}\\
\mbox{ENTSO-E} generation per type & Generation & MW & 15\,min to hourly & Country & Net electricity production time series  & \cite{2020ENTSO-EType} \\
Eurostat energy balance & Generation & ktoe & yearly & Country  & Gross electricity production and associated energy input & \cite{EnergyEurostat} \\
EU ETS Data (EUTL) & Emissions & t CO$_2$ & yearly & Power plant  & Reported emissions in the EU ETS mechanism &  \cite{EUROPALog}\\
UNFCCC & Emissions & t CO$_2$ & yearly & Country & Reported emissions for different sectors in all countries to the UNFCCC & \cite{NationalAgency} \\ \bottomrule
\end{tabular}
\caption{Overview of used datasets and their properties}
\label{tab:used_data}
\end{table*}

\subsection{ENTSO-E per generation unit data}
\label{sec:entsoe_unit}
The dataset ``Actual Generation per Generation Unit'' provided by~\mbox{ENTSO-E} contains dispatch time series for large power plant units located in the \mbox{ENTSO-E} area~\cite{2020ENTSO-EUnit}. The generation data is published five days after the operational period. Since this dataset has only been established in 2015, up to now, no critical review is available in the literature. The time series has been corrected for obviously erroneous entries and duplicates, but we did not fill gaps due to the heterogeneous and often irregular nature of the individual dispatch time series. In the final step, all generation time series were converted to the same temporal resolution of one hour.

\subsection{ENTSO-E Statistical Factsheets}
\label{sec:entsoe_sheet}
The annually published ``Statistical Factsheet'' from~\mbox{ENTSO-E} contains, among other things, yearly aggregated values for electricity load and generation for European countries~\cite{ENTSO-E2018}. In the fact sheet, it is stated that the data has been consolidated by taking into account national statistical resources, but no further details about the correction process are given. 

\subsection{\mbox{ENTSO-E} per production type data}
\label{sec:entsoe_prod}
The dataset ``Actual Generation per Production Type'' provided by \mbox{ENTSO-E} contains power generation time series for all countries in the \mbox{ENTSO-E} area for different production types~\cite{2020ENTSO-EType}. The temporal resolution of the data ranges, depending on the country, from 15 minutes to up to 1 hour. The completeness and consistency of the data varies across countries and across generation types~\cite{Hirth2018}. For instance, for Germany, the aggregated generation time series from \mbox{ENTSO-E} for gas only covers around 50\,\% of the corresponding generation reported by the German Working Group on Energy Balances~\cite{AGEnergiebilanzene.V.2020Stromerzeugung2020}. We have cleaned the data and checked for gaps and duplicates. Gaps with a length of up to two hours long were filled linearly, and duplicates were removed. For Sweden, further manual corrections were implemented, because they showed an incorrect allocation in the generation technologies. A detailed description of all corrections can be found in the published code python \cite{co2emissionsfactorsgithub}. In the final step, all generation time series were converted to the same temporal resolution of one hour. Although \mbox{ENTSO-E} classifies more than 20 different generation types, we have reduced the technologies to twelve types, for reasons of applicability. The corresponding assignment of the \mbox{ENTSO-E} generation technologies to the selected technology is shown in Tab.~\ref{tab:technologies}.

\begin{table}[h]
\begin{tabular}{p{2.2cm}p{5.5cm}}
\toprule
Technology & Subsumed ENTSO-E technologies \\
\midrule
hard\_coal & Fossil hard coal \\
lignite & Fossil brown coal/lignite \\
gas & Fossil gas \\
other\_fossil & Fossil coal-derived gas, fossil peat, other, fossil oil, fossil oil shale \\
nuclear & Nuclear \\
biomass & Biomass \\
waste & Waste \\
other\_renewable & Geothermal, marine, other renewable \\
hydro & Hydro pumped storage, hydro run-of-river and poundage, hydro water reservoir \\
solar & Solar \\
wind\_onshore & Wind onshore \\
wind\_offshore & Wind offshore \\ \bottomrule
\end{tabular}
\caption{Mapping for generation technologies from \mbox{ENTSO-E} to the classification used in this study.}
\label{tab:technologies}
\end{table}

\subsection{Eurostat energy balances}
\label{sec:balance}
The energy balances published by Eurostat provide an overview of energy products and their flow in the economy~\cite{EnergyEurostat}. This accounting framework for energy products allows studying the total amount of energy extracted from the environment, traded, transformed and used by the different European countries. In the balance sheet,  also the relative contribution of each energy carrier (fuel, product) is represented. For the method presented in this study, we incorporate data for energy input quantities used for power generation, as well as the resulting gross electricity generation. In contrast to the data provided by \mbox{ENTSO-E}, Eurostat subdivides the energy transformation input into different energy carriers. For example, \mbox{ENTSO-E} distinguishes between lignite and hard coal, whereas Eurostat lists nine different coal types (anthracite, coking coal, other bituminous coal, sub-bituminous coal, lignite, coke oven coke, gas coke, coal tar, brown coal briquettes) in their energy balance. The resulting electricity output is given by Eurostat as the gross electricity production, whereas \mbox{ENTSO-E} reports net production, i.e., subtracting the power plant's own consumption.

\subsection{EU emissions trading system}
\label{sec:euets}
For determining emissions at the power plant level, we use data from the EU~ETS~\cite{EUAction}. For the electricity sector, the EU~ETS register records all emissions from enlisted individual power plants. This central information is represented by the verified emissions, which refers to the number of CO$_2$ certificates (amount of emitted CO$_2$) needed by an installation during the review period. For our method, the so-called free allocations (certificates distributed to installations free of charge) are also taken into account. Free certificates are granted to installations when they provide products that are transferred to a sector that is not covered by the EU~ETS system, such as heat for private households \cite{EuropeanCommission2015EUHandbook}.

\subsection{National emissions reported to the UNFCCC}
\label{sec:unfccc}
The UNFCCC and the Kyoto Protocol 2020 specify and regulate emissions reporting for all participating countries. The corresponding annual reports contain data for all emissions in individual countries and sectors. The accuracy of the emission data depends on the method used in the respective countries. It should be noted that the reporting guidelines allow some degree of freedom for the individual countries with respect to the application of different methods for calculating emissions attributed to different sectors. Generally, the corresponding reporting schemes are biased toward overestimating emissions if less elaborated calculation methodologies are applied \cite{Eggleston20062006Inventories}. The individual reports provide information about the underlying method for each country data. For the method presented in this study, yearly emissions per country from the public electricity and heat sector are used.

\section{Methodology}
\label{sec:method}

In the following, we present two methodological approaches to assess the carbon emission intensity of electricity generation in European countries. First, the bottom-up method based on emission data for individual power plants and time series for electricity generation is introduced in Section~\ref{sec:bottomupmethod}. The top-down method, which uses energy balances and reported per country annual emissions is reviewed in Section~\ref{sec:topdownmethod}.

\subsection{Bottom-up method}
\label{sec:bottomupmethod}
\begin{figure}[h]
	\centering
	\includegraphics[width=0.4\textwidth]{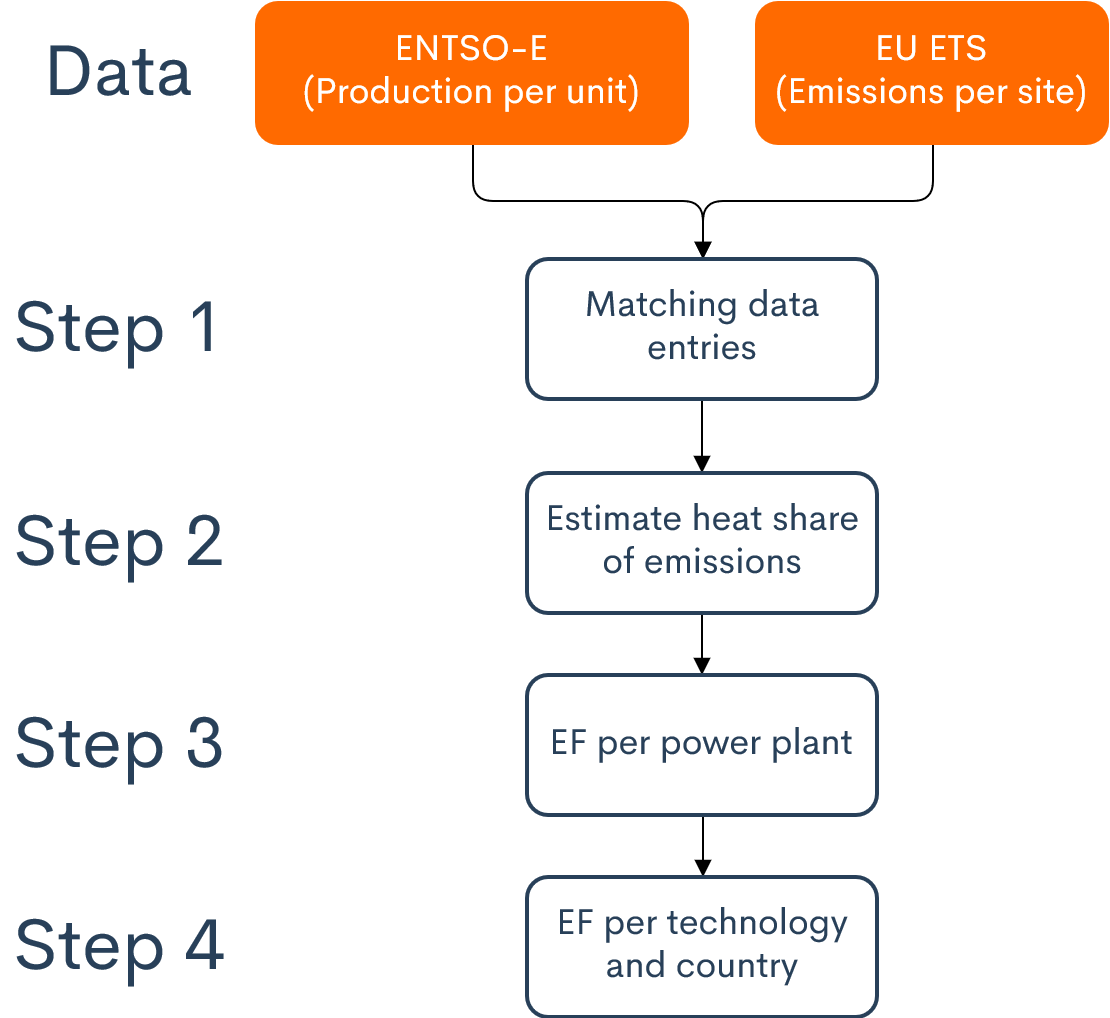}
	\caption{Process illustration of the bottom-up method}
	\label{fig:Bottom-up_process}
\end{figure}

The key outcome of the bottom-up method are yearly aggregated CO$_2$ EFs of electricity generation per type and country. It is based on individual emission and generation data per power plant. The method consists of four main steps involving two data sources, as visualized in Figure~\ref{fig:Bottom-up_process}. In the first step, two datasets are matched, one containing production time series (\mbox{ENTSO-E} generation per unit), and one stating yearly CO$_2$ emissions reported as part of the EU~ETS scheme (cp. Section~\ref{matching}). In a second step, the share of CO$_2$ emissions, which can be allocated to heat generation, as further specified in Section~\ref{heat}, is estimated. In a subsequent step, the EFs per power plant are calculated based on the emission and production data from the matched datasets, as explained in Section~\ref{EF_pp}. In the final step, a representative power plant sample for each technology and country is chosen to determine their respective EFs (cp. Section~\ref{EF_ct}).

The method can be applied for each country separately. For convenience, we omit an explicit index for the country in our presentation. If a temporal index $t$ is given, the quantity is defined for the hour $t$, whereas quantities without a temporal index are defined for the year under consideration. If not otherwise stated, all data discussed in the following refers to the year 2018.

\subsubsection{Matching records in \mbox{ENTSO-E} and EUTL datasets}\label{matching}
To derive individual EFs for power plants, we combine two datasets. Hourly production time series for large power plants are published by \mbox{ENTSO-E} through the transparency portal~\cite{2020ENTSO-EUnit}. In the following, $G_{t}(\alpha)$ denotes the per unit generation for the power plant with index $\alpha$ in hour $t$. Carbon dioxide emission data for power plants is published through reports from the European Union Transaction Log (EUTL), which contains all transactions between accounts from the EU~ETS mechanism~\cite{EUROPALog}. This data yields the emissions $CO_2(\alpha)$ associated with electricity generation from power plant $\alpha$ for the given year. To associate the annual generation $G(\alpha)=\sum_{t}G_{t}(\alpha)$ with the corresponding emissions $CO_{2}(\alpha)$, we matched the \mbox{ENTSO-E} energy identification codes (EIC) with the ETS plant identifier (EUTL-ID). This matching procedure is non-trivial, since entries from both datasets use different name formats or even different names for the same power plants. Using other power plant databases along with manual searches resulted in a matching list containing 859 entries with a total installed capacity of 298.07\,GW  (233 coal units with 87.13\,GW, 452 gas units with 152.77\,GW and 131 lignite units with 45.17\,GW) \cite{gemwiki,JRCPPDB,Gotzens2019,OPSD2020}. The matching between the two datasets leads to certain data inaccuracies. First, it is impossible to match all generation units from the \mbox{ENTSO-E} dataset to an installation listed in the EUTL dataset. In total, 919 units out of 1759 generation units are unmatched, representing 257.02\,GW of installed capacity. Additionally, several individual generation units in the \mbox{ENTSO-E} dataset are listed under a single location name in the EUTL. In these cases, we allocated the emissions to the units according to the installed capacities. A detailed description and evaluation of all matched power plants is available via the data export within the python code \cite{co2emissionsfactorsgithub,unnewehr_jan_frederick_2021_5603077}.

\subsubsection{Determining the heat generation}\label{heat}
The reported emissions $CO_{2}(\alpha)$ do not discriminate between emissions associated with electricity or heat generation. To take the heat extraction from power plants into account when calculating EFs, the amount of CO$_2$ which can be allocated to the heat export must be estimated. Here, we approximate emissions that can be allocated to heat production using the emission allowances allocated free of charge in the EU~ETS system. Based on the allocation quantity in 2018, a free allocation of 50\,\% of the emission allowances for heat production is assumed, which is consistent with~\cite{EuropeanCommission2015EUHandbook}. For 2018, subtracting twice the free allowances from the reported emission thus yields the annual emissions ${CO_{2}}^{\mathrm{el}}(\alpha)$ which are associated with the electricity generation from the power plant~$\alpha$.

\subsubsection{Emission factor per power plant}\label{EF_pp}

The annual average emission factor $EF(\alpha)$ for a power plant~$\alpha$ is calculated as the ratio of the total generation per year and power plant, $G(\alpha)=\sum_t G_{t}(\alpha)$, to the corresponding emissions per power plant, ${CO_{2}}^{\mathrm{el}}(\alpha)$,

\begin{equation}\label{EF_y}
EF(\alpha)= \frac{{CO_2}^{\mathrm{el}}(\alpha)}
{G(\alpha)}~.
\end{equation}

After calculating the individual values $EF(\alpha)$, a plausibility check was performed for each power plant. The EF was included only in the final list if it was within the plausibility range, based on EFs calculated by the National Renewable Energy Laboratory (NREL)~\cite{NREL2021}. They performed a systematic review of about 3\,000 publications for life cycle assessment of electricity generation technologies. The resulting plausibility range for our chosen technologies is based on 110 studies with 268 estimates for EFs and shown in Table~\ref{tab:EF_plausibility}. If the calculated power plant EF is outside this plausibility range, we assume potential issues with the underlying data (errors, gaps, misreporting) and omit the corresponding power plant from further calculations.

\begin{table}[]
\begin{tabular}{@{}p{2cm}p{1.8cm}p{2cm}p{1.8cm}@{}}
\toprule
& \multicolumn{3}{c}{Emissions} \\
Fuel type & Minimum {(}gCO$_2$/kWh{)} & Median {(}gCO$_2$/kWh{)} & Maximum {(}gCO$_2$/kWh{)} \\ \midrule
Gas      & 300             & 500                  & 1\,000              \\ 
Coal / Lignite     & 675             & 1\,000                  & 1\,700             \\ 
Other fossil    & 300             & 850                 & 1\,700             \\
\bottomrule
\end{tabular}
\caption{Upper and lower limits for plausibility check of the emission factor calculation; the values are taken from literature review done by the National Renewable Energy Laboratory (NREL)~\cite{NREL2021}.}
\label{tab:EF_plausibility}
\end{table}

\begin{figure}[h]
	\centering
	\includegraphics[width=0.45\textwidth]{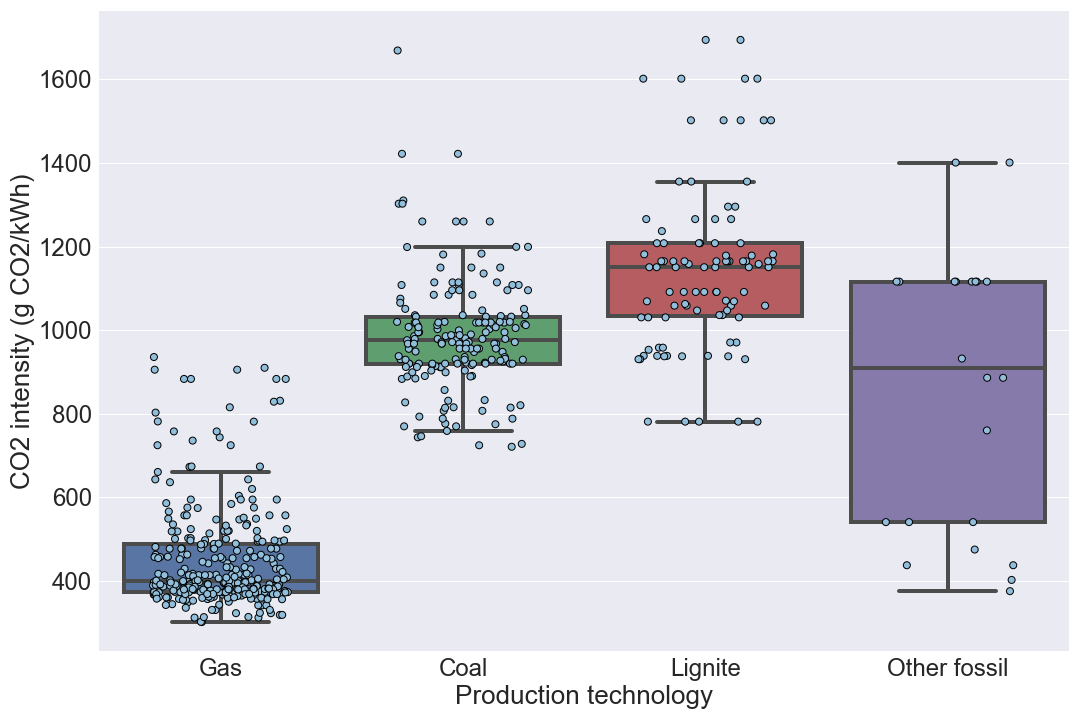}
	\caption{Average emission factor per technology and distribution of per unit $EF(\alpha)$ over the entire dataset. The points represent the annual average emission factor $EF(\alpha)$ for an individual power plant $\alpha$. The bars represent the average emission factor per technology in the entire dataset.}
	\label{fig:CO2_technology}
\end{figure}

As a result, 595 European generation units and their individual average annual EFs were considered valid. Based on the \mbox{ENTSO-E} time series, these power plants supplied 717\,TWh of electricity to the grid in Europe, which represents roughly 50\,\% of the conventional generation in 2018 reported in \cite{ENTSO-E2018}. The resulting range of EFs per technology is shown in Figure~\ref{fig:CO2_technology}. For gas, coal and lignite, it is visible that most of the entries are distributed around the average value. For the few nonspecific power plants (other fossil), the variance of the calculated EF is considerably higher. This indicates vaious underlying technologies, including for instance oil or waste power plants.

\subsubsection{Emission factors per country and technology}\label{EF_ct}

For each country, we partition the set of power plants $\mathcal{G}$ into four sets $\mathcal{G}(\mathrm{tech})$ corresponding to the different technology categories. To obtain country-specific EFs per technology, the ratio of all emissions associated with electricity generation from a certain technology class to the corresponding generation is calculated:

\begin{equation}\label{EF_i_tech}
EF(\mathrm{tech})=\frac{
\sum_{\alpha\,\in\,\mathcal{G}(\mathrm{tech})}{CO_{2}}^{\mathrm{el}}(\alpha)
}
{
\sum_{\alpha\,\in\,\mathcal{G}(\mathrm{tech})}G(\alpha)
}
\end{equation}

Note that we calculate these per technology EFs here for countries, but since the underlying quantities are given for individual power plants with a known geographic location, analogous calculations can be performed on other spatial scales as well.

\begin{figure}[h]
	\centering
	\includegraphics[width=0.45\textwidth]{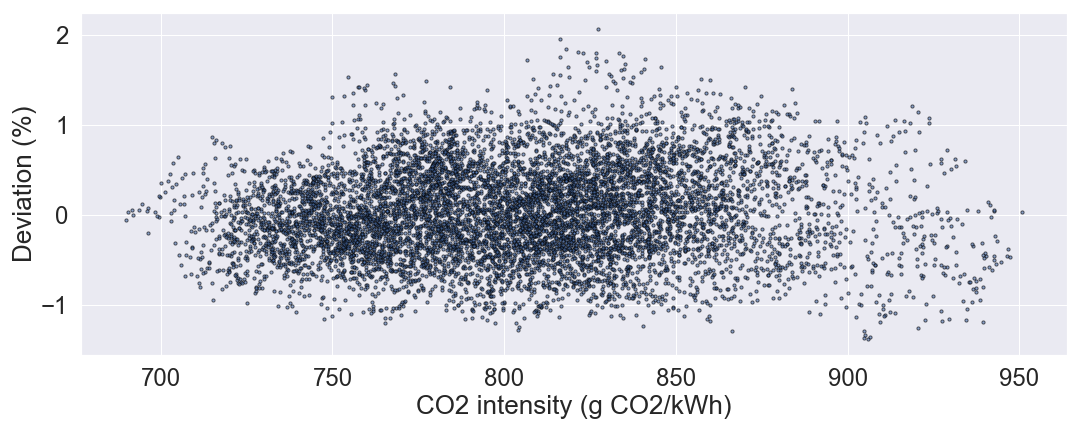}
	\caption{Relative deviation $\left(\tilde{CI}^{\mathrm{tech}}-\tilde{CI}^{\mathrm{plant}}\right)/\tilde{CI}^{\mathrm{tech}}$ between the hourly carbon intensity based on unit specific  and technology specific emission factors, respectively. Each dot corresponds to a value for one hour, with the deviation on the y-axis and the technology-based carbon intensity $\tilde{CI}^{\mathrm{tech}}$ on the x-axis. The carbon intensities have been calculated over all power plants from the validated matched dataset.}
	\label{fig:deviation_CO2_unit_tech}
\end{figure}

Both emission factors $EF(\alpha)$ and $EF(\mathrm{tech})$ are calculated as yearly averages. To calculate an emission signal for the generation mix in an individual hour, these factors must be multiplied with the corresponding hourly generation. Since the time series for the generation per production type can be shown to have higher coverage,  we use per technology emission factors $EF(\mathrm{tech}$) in the following to calculate such hourly emission signals as well as the average carbon intensity per country. However, pooling generation capacities from the same technology category will smoothen the heterogeneity in the emission intensity and dispatch of individual power plants. To estimate this loss of information, we compare the hourly carbon intensity per country associated with the power plant dispatch time series $G_{t}(\alpha)$ and emission factors $EF(\alpha)$ on the one hand, and $EF(\mathrm{tech})$ on the other hand:

\begin{align}
\tilde{CI}^{\mathrm{plant}}_t &= \frac{\sum_{\alpha}G_{t}(\alpha)EF(\alpha)
}{\sum_{\alpha}G(\alpha)
}~,
\\
\tilde{CI}^{\mathrm{tech}}_t &=  \frac{\sum_{\mathrm{tech}}G_t(\mathrm{tech})EF(\mathrm{tech})
}{
\sum_{\mathrm{tech}}G_t(\mathrm{tech})
}~.
\end{align}
Here, we use
\begin{equation}
G_t(\mathrm{tech})=\sum_{\alpha\,\in\,\mathcal{G}(\mathrm{tech})}G_t(\alpha)~.
\end{equation}
Figure~\ref{fig:deviation_CO2_unit_tech} shows the relative deviation $\left(\tilde{CI}^{\mathrm{tech}}-\tilde{CI}^{\mathrm{plant}}\right)/\tilde{CI}^{\mathrm{tech}}$ for the resulting carbon intensity in gCO$_2$/kWh when using technology-specific EFs versus the original unit-specific EFs. One point in this figure represents the resulting carbon intensity for one hour in the year 2018, based on the generation and emissions of the power plants in the validated dataset. Note that this calculation was done based on per unit generation time series from the matched data, so the resulting values do not correspond to the  carbon intensity of the entire generation mix since, among other things, renewable generation is missing. The results shown in Fig.~\ref{fig:deviation_CO2_unit_tech} indicate that for the given dataset, the deviations range between minus 2\,\% and plus 1.5\,\%, without a clear bias for over- or underestimation. Therefore, we conclude that the loss of information originating from using per production type emission factors $EF(\mathrm{tech})$ compared to per power plant emission factors $EF(\alpha)$ is negligible for calculating the per country hourly carbon intensity of electricity generation.

The per technology emission factors $EF(\mathrm{tech})$ can be used along hourly per technology generation time series to provide an hourly carbon intensity signal. Such time series are given by \mbox{ENTSO-E} per country, denoted in the following by $G^{\mathrm{type}}_t(\mathrm{tech})$. This dataset has a higher coverage compared to the per unit generation time series, of which, only a subset is included in the calculation of the per power plant emission factors $EF(\alpha)$ due to the matching procedure. However, for most countries and technologies, the aggregated generation time series $\sum_t G^{\mathrm{type}}_t(\mathrm{tech})$ yields a lower annual generation than the consolidated values reported by \mbox{ENTSO-E} in the Statistical Factsheet.To implement a first-order correction of the time series, we thus scaled the hourly generation to yield the same reported annual aggregate per country and technology. If the aggregated time series was larger than the corresponding value in the Factsheet, we kept these generation values unchanged.

To assess the representativeness of the per technology emission factors $EF(\mathrm{tech})$, we compared the aggregated underlying per power plant generation with the corresponding aggregated (scaled) per technology generation data,
\begin{equation}\label{eq:coverage}
    \mathrm{Coverage(tech)}=\frac{
    \sum_t\sum_{\alpha\,\in\,\mathcal{G}(\mathrm{tech})}G_t(\alpha)
    }
    {\sum_t G_t^{\mathrm{type}}(\mathrm{tech})}~.
\end{equation}

Table~\ref{tab:EF_per_country_tech} displays the calculated emission factors $EF(\mathrm{tech})$ per technology for each country, and gives an overview of the underlying data. We assume an EF to be representative for a country and a technology if the corresponding coverage as calculated in Eq.~(\ref{eq:coverage}) is larger than $25\,\%$. In Table~\ref{tab:EF_per_country_tech}, we list for each country and technology the number of power plants and their aggregated capacity as contained in the \mbox{ENTSO-E} per unit generation time series dataset, the subset which we have matched to power plants from the EUTL, and the final validated matched dataset, which fulfills the coverage criterion of $25\,\%$. 
It should be emphasized that the coverage of the matched power plants can be extended in the future, facilitated by the publication of all code as well as the matching used for the results in this study, and the public availability of all underlying data.

\begin{table*}[]
\begin{tabular}{@{}llrrrrrrrr@{}}
\toprule
& & & & \multicolumn{6}{c}{Data et analysis} \\
Country & Technology & EF {(}gCO$_2$/kWh{)} & Coverage {(}\%{)} & \multicolumn{2}{c}{Plausible} & \multicolumn{2}{c}{Matched} & \multicolumn{2}{c}{ENTSO-E} \\ 
& & &  & Cap. {(}GW{)} & Count & Cap. {(}GW{)} & Count & Cap. {(}GW{)} & Count \\
\midrule
AT & gas & 398.29 & 24.0 & 2384.014 & 7 & 7457.226 & 12 & 7457.226 & 12 \\
AT & hard\_coal & 884.07 & 39.0 & 693.857 & 1 & 1439.148 & 3 & 1439.148 & 3 \\
BE & gas & 389.64 & 64.0 & 14216.707 & 17 & 16372.357 & 19 & 17181.875 & 20 \\
CZ & gas & 780.84 & 39.0 & 1729.036 & 1 & 1729.036 & 1 & 1729.036 & 1 \\
CZ & hard\_coal & 985.55 & 36.0 & 1243.426 & 3 & 1243.426 & 4 & 1243.426 & 4 \\
CZ & lignite & 928.3 & 58.0 & 19775.004 & 18 & 26607.733 & 25 & 21876.551 & 21 \\
DE & gas & 354.6 & 17.0 & 15235.146 & 29 & 29445.917 & 58 & 29024.315 & 53 \\
DE & hard\_coal & 934.14 & 94.0 & 68655.321 & 42 & 78632.75 & 51 & 73274.329 & 48 \\
DE & lignite & 1125.56 & 96.0 & 129230.488 & 34 & 129230.488 & 35 & 129230.488 & 35 \\
DE & other\_fossil & 1401.1 & 30.0 & 1971.58 & 2 & 3759.405 & 4 & 6423.292 & 11 \\
DK & gas & 329.78 & 10.0 & 273.443 & 1 & 273.443 & 1 & 273.443 & 1 \\
DK & hard\_coal & 775.84 & 69.0 & 4792.089 & 6 & 6332.337 & 8 & 5562.213 & 7 \\
EE & other\_fossil & 1057.97 & 109.0 & 9592.431 & 11 & 9592.431 & 13 & 9592.431 & 13 \\
ES & gas & 390.43 & 49.0 & 26209.431 & 63 & 26795.465 & 72 & 26795.465 & 72 \\
ES & hard\_coal & 1044.87 & 99.0 & 32138.25 & 19 & 35004.169 & 26 & 35004.169 & 26 \\
FI & hard\_coal & 754.59 & 47.0 & 2816.937 & 4 & 5099.777 & 12 & 5099.777 & 12 \\
FI & other\_fossil & 759.81 & 13.0 & 725.116 & 1 & 725.116 & 1 & 3397.165 & 10 \\
FR & gas & 396.98 & 49.0 & 15259.9 & 11 & 19720.563 & 18 & 20270.423 & 19 \\
FR & hard\_coal & 928.83 & 73.0 & 4256.081 & 4 & 5651.195 & 5 & 5651.195 & 5 \\
GB & gas & 488.64 & 64.0 & 83317.83 & 57 & 88668.614 & 77 & 91207.961 & 84 \\
GB & hard\_coal & 1125.0 & 46.0 & 7652.145 & 19 & 11004.849 & 25 & 11004.849 & 27 \\
GR & gas & 457.38 & 65.0 & 9928.104 & 10 & 15146.233 & 14 & 15146.233 & 14 \\
GR & lignite & 1429.22 & 100.0 & 15188.996 & 14 & 15188.996 & 14 & 15188.996 & 14 \\
HU & gas & 379.65 & 65.0 & 4081.735 & 9 & 4497.068 & 10 & 4497.068 & 10 \\
HU & lignite & 1355.62 & 81.0 & 3850.49 & 3 & 3850.49 & 3 & 3850.49 & 3 \\
IE & gas & 392.88 & 44.0 & 6552.647 & 7 & 12047.551 & 16 & 12047.551 & 16 \\
IE & hard\_coal & 1032.2 & 89.0 & 1894.775 & 3 & 1894.775 & 3 & 1894.775 & 3 \\
IT & gas & 395.17 & 52.0 & 64447.804 & 53 & 72774.603 & 69 & 73504.645 & 72 \\
IT & hard\_coal & 997.67 & 59.0 & 17039.711 & 15 & 25848.683 & 18 & 25848.683 & 18 \\
IT & other\_fossil & 458.37 & 22.0 & 13726.939 & 8 & 23804.91 & 20 & 36286.053 & 43 \\
NL & gas & 393.24 & 40.0 & 28718.568 & 19 & 36881.05 & 34 & 38298.322 & 37 \\
NL & hard\_coal & 1183.07 & 24.0 & 4080.723 & 1 & 27130.255 & 6 & 27130.255 & 6 \\
PL & gas & 370.31 & 10.0 & 1294.905 & 2 & 2216.311 & 3 & 7245.647 & 5 \\
PL & hard\_coal & 942.41 & 56.0 & 44401.449 & 52 & 45302.785 & 56 & 64890.861 & 82 \\
PL & lignite & 1158.65 & 78.0 & 35099.139 & 13 & 41691.758 & 19 & 45574.123 & 26 \\
PT & gas & 430.33 & 42.0 & 6169.294 & 7 & 10187.982 & 10 & 10187.982 & 10 \\
RO & gas & 426.42 & 28.0 & 2991.592 & 9 & 6987.634 & 21 & 6987.634 & 21 \\
RO & hard\_coal & 1157.44 & 72.0 & 852.205 & 4 & 852.205 & 6 & 852.205 & 6 \\
RO & lignite & 1000.76 & 88.0 & 12869.138 & 11 & 12869.138 & 14 & 12869.138 & 14 \\
SE & gas & 361.8 & 68.0 & 408.286 & 1 & 408.286 & 1 & 408.286 & 1 \\
SK & hard\_coal & 903.09 & 65.0 & 782.936 & 2 & 782.936 & 2 & 782.936 & 2 \\
SK & lignite & 1295.55 & 87.0 & 1167.766 & 2 & 1167.766 & 3 & 1167.766 & 3 \\ \bottomrule
Sum &  &  &  & 717715.44 & 595 & 866316.86 & 812 & 907398.43 & 890 \\ \bottomrule
\end{tabular}
\caption{Technology-specific emission factors $EF(\mathrm{tech})$ per country determined via the bottom-up method. The coverage of the underlying per unit generation time series from matched power plants is defined in Eq.~(\ref{eq:coverage}). A coverage value of more than $100\,\%$ for EE is due to misreported values in the \mbox{ENTSO-E} Statistical Factsheet. Total capacity per country and number of generation units is given for the entries of the original per unit generation dataset, the matched entries, and the validated matched entries selected through the plausibility criterion shown in Table~\ref{tab:EF_plausibility}.}

\label{tab:EF_per_country_tech}
\end{table*}

\subsubsection{Average emission intensity per country}

The carbon intensity per country is calculated as
\begin{equation}
    CI=\frac{\sum_{\mathrm{tech}}\sum_t \left(G^{\mathrm{type}}_t(\mathrm{tech})\cdot EF(\mathrm{tech})\right)}
    {\sum_t\sum_{\mathrm{tech}}G^{\mathrm{type}}_t(\mathrm{tech})}~,
\end{equation}
using the scaled per generation type time series $G^{\mathrm{type}}_t(\mathrm{tech})$ and the per technology emission factors $EF(\mathrm{tech})$ as calculated in the last section. If no representative EF is available for one country and technology pair, we use a weighted average value for this technology, based on the EFs given for the other countries ($1\,184.81$, $974.91$, $419.79$, $919.32$ gCO$_2$/kWh for lignite, hard coal, gas, and other fossil). The resulting CI for each county is displayed in Table~\ref{tab:CI_countries}, where these results are compared with corresponding values calculated using the top-down method presented in the next section.

\subsection{Top-down method}
\label{sec:topdownmethod}
\begin{figure}[h]
	\centering
	\includegraphics[width=0.4\textwidth]{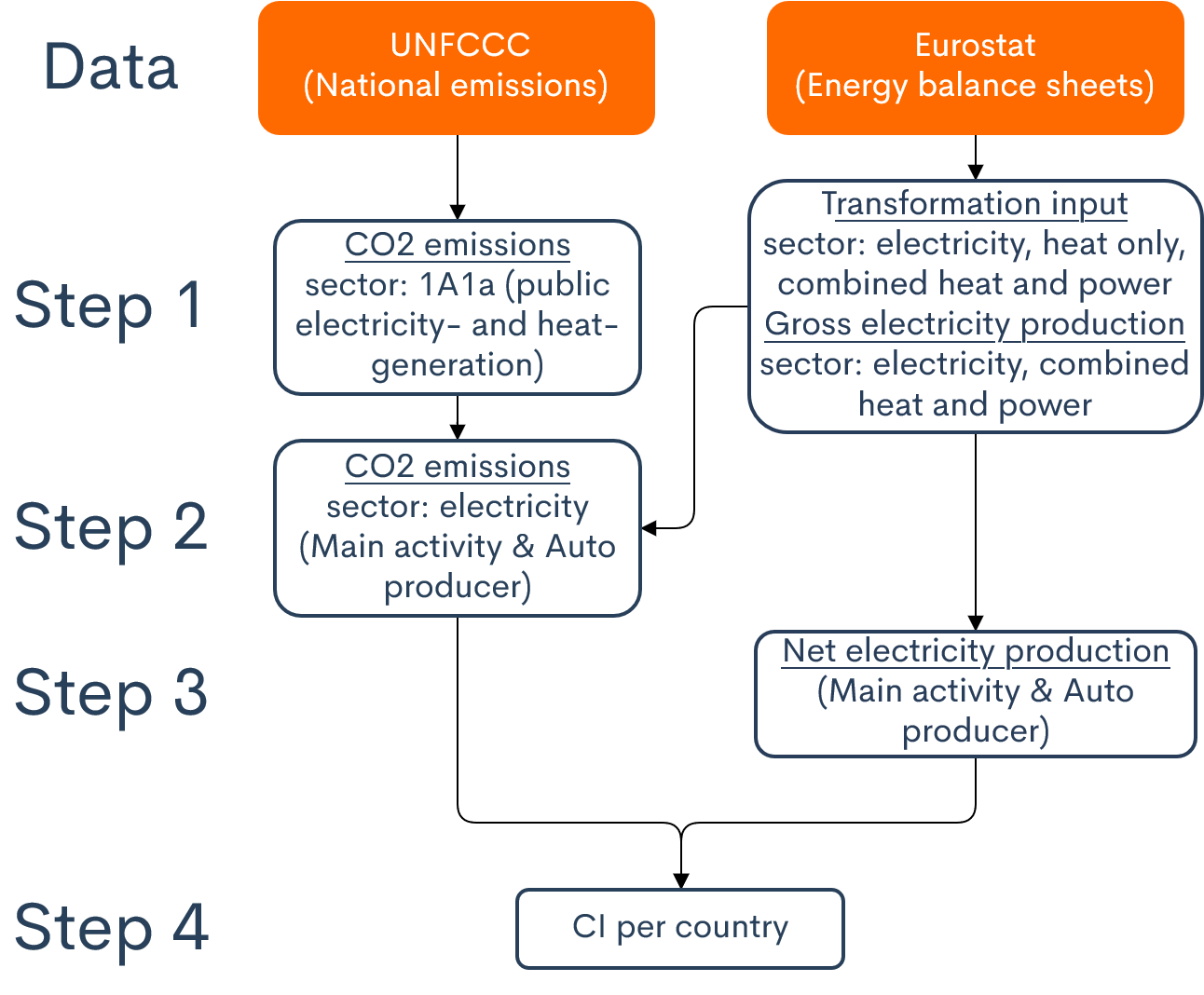}
	\caption{Process illustration of the top-down method}
	\label{fig:Top-down_process}
\end{figure}

The top-down method uses nationally reported data on electricity and heat generation as well as associated emissions. This approach is applied by the European Environment Agency (EEA), which publishes average annual values for the carbon dioxide intensity of electricity generation on country- and EU-level \cite{EEA2021}. The input datasets are the energy balance sheets published by Eurostat~\cite{EnergyEurostat} for the electricity and heat generation, and the reported emissions to the UNFCCC \cite{NationalAgency} for the associated emissions. The derivation of the national carbon intensity factors involves four steps which are visualized in Figure~\ref{fig:Top-down_process}. The input data is always given for a specific country and year. Thus, for convenience, we omit corresponding indices in the notation used in the following description of the process steps.

\subsubsection{CO2 emissions and energy balance}

The input values $CO_2(MAP)$ for the emissions per country and year are given by the total emissions from main activity produces (MAP) with activities public electricity generation, public combined heat and power generation, and public heat generation, as reported to the UNFCC (sector 1.A.1.a). These reported values exclude life-cycle greenhouse gas emissions, and therefore assume no CO$_2$ emissions from nuclear renewable power generation. Also, biomass-related emissions are not reported in the energy sector according to the UNFCC Reporting Guidelines, but they are associated with the Land Use, Land Use Change and Forestry (LULUCF) sector. Accordingly, corresponding emissions are not included in the input data here~\cite{EEA2021}.

In order to associate the reported emissions $CO_2(MAP)$ with electricity and heat generation, input data from the energy balance sheet \emph{nrg\_bal\_c}, published by Eurostat, were used. The main input variables are the energy transformation input values $TI(p,s)$ for main activity producers ($p=MAP$) and auto producers ($p=AP$), distinguished by the sectors electricity only ($s=E$), heat only ($s=H$) and combined heat and power ($s=CHP$). Main activity producers generate electricity and heat for sale to third parties as their primary activity, whereas auto producers generate electricity and heat for their own use as a supporting activity for their main activity. The input values $TI(p,s)$ have been aggregated over the fuel types solid fossil fuel, oil and petroleum products (excl. biofuel), natural gas, manufactured gases, peat and peat products, oil shale and oil sands, and non-renewable waste. Additionally, the resulting gross electricity production $GE(p,s)$ for $p=AP,MAP$ and $s=E,CHP$, and the derived heat $dh(p,s)$ for $p=AP,MAP$ and $s=H,CHP$ enter the calculation.

\subsubsection{CO$_2$ emission intensity calculation}
The CO$_2$ emission intensity of electricity generation $CI$ is calculated from the ratio of all CO$_2$ emissions from electricity generation to total electricity generation. For the given input data, this apparently simple definition involves certain methodological choices:
\begin{itemize}
    \item Auto producers: The reported emissions $CO_2(MAP)$ exclude emissions from auto producers, so the calculation of the carbon intensity $CI$ has to be limited to main activity producers only, or the auto producers' emissions must be estimated based on their share in the energy balance. We follow the latter approach and include auto producers in the calculation of the carbon intensity $CI$.
    \item Heat generation: The reported emissions $CO_2(MAP)$ include emissions from heat generation. Consequently, the corresponding share of emissions must be estimated based on the energy balance and then be subtracted, which in particular involves some approximations done with respect to the combined heat and power sector.
\end{itemize}

\paragraph{Main activity producers} 
As stated above, the reported emissions $CO_2(MAP)$ represent not only emissions associated with electricity generation but also consider public heat generation. The share of emissions associated with electricity generation is estimated based on the corresponding share of energy transformation input into the energy balance:
\begin{equation}\label{eq:CO2_MAP}
{CO_2}^{\mathrm{el}}(MAP)=CO_2(MAP)\cdot\frac{TI^{\mathrm{el}}(MAP)}{TI(MAP)}~,
\end{equation}
with
\begin{align}
\label{eq:TI_def_1}
    TI(MAP) &=\sum_{s=E,CHP,H}TI(MAP,s)~,\\
\label{eq:TI_def_2}
    TI^{\mathrm{el}}(MAP) &=TI(MAP,E)\nonumber\\
    &\phantom{=}+TI(MAP,CHP)-\frac{dh(MAP,CHP)}{0.9}~.
\end{align}

The denominator in Eq.~(\ref{eq:CO2_MAP}) contains the sum of the transformation input from the three relevant sectors for main activity producers as calculated in Eq.~(\ref{eq:TI_def_1}). The enumerator contains the transformation input from the electricity and combined heat and power generation. In a later step, the CO$_2$ emissions that can be attributed to the heat generation must be subtracted. These are estimated from the derived heat of the sector, assuming a typical boiler efficiency of $90\,\%$ for heat production \cite{Li2009NOxStaging}. Note that the EEA uses a slightly different variant of Eq.~(\ref{eq:TI_def_2}), including the additional term $TI(MAP,H)-dh(MAP,H)/0.9$ in the enumerator. This term first adds the transformation input for heat only generation, and then subtracts the corresponding value based on the derived heat from this sector, also with an assumed efficiency of $90\,\%$. Since both approaches yield similar results, we use the simpler version displayed in Eq.~(\ref{eq:TI_def_2}).
\paragraph{Auto producers}
Since the reported emissions to the UNFCC exclude emissions for auto producers, these emissions are estimated based on the ratio between the electricity-related energy transformation input for auto producers and for main activity producers, respectively:
\begin{equation}
\frac{{CO_2}^{\mathrm{el}}(AP)}{{CO_{2}}^{\mathrm{el}}(MAP)}
=
\frac{TI^{\mathrm{el}}(AP)}{TI^{\mathrm{el}}(MAP)}~.
\end{equation}
The energy transformation input $TI^{\mathrm{el}}(AP)$ for auto producers is calculated analogously to the one for main activity producers in Eq.~(\ref{eq:TI_def_2}).

\paragraph{Carbon emission intensity}
To derive the carbon emission intensity of electricity generation, the aggregated  emissions $CO_2(MAP)$ and $CO_2(AP)$ from the main activity and auto producers must be divided by the corresponding electricity generation. Following the approach by the EEA, this generation is given by the sum of gross electricity production $GE(p,s)$ for $p=MAP,AP$ and $s=E,CHP$ from the Eurostat energy balance sheet. Additionally, we transform the gross electricity production to the net electricity production assuming a self-consumption of 6\,\% \cite{AGEnergiebilanzene.V.2020Stromerzeugung2020} for all power plants:
\begin{equation}
    CI=
    \frac{{CO_2}^{\mathrm{el}}(MAP)+{CO_2}^{\mathrm{el}}(AP)}
    {\sum_{p=MAP,AP}\sum_{s=E,CHP}GE(p,s)\cdot 0.94}~.
\end{equation}

\section{Results}
\label{sec:application}

In the following, we present the results of the application of the top-down and bottom-up method on the emission and generation datasets for the year 2018.

\subsection{Carbon intensity of countries}
\label{sec:country_CI}
Table~\ref{tab:CI_countries} shows the CI per country for the year 2018 based on the top-down and the bottom-up method, respectively. Both methods yield comparatively low CIs for Sweden, Norway and France due to their high shares of nuclear and hydro power generation. At the other end of the scale, Greece, Cyprus, Poland and Estonia have high CIs associated with their fossil-fuel based generation mix~\cite{ENTSO-E2018}. The table shows that although the top-down and bottom-up methods in general yield similar results, for some countries, significant differences occur. There is no overall bias for over- or underestimation of one method over the other, but variations differ in magnitude and sign. The CI of Lithuania, for instance, is $53\,\%$ higher with the bottom-up method compared to the top-down approach, whereas for Norway, it is $19\,\%$ lower. Other literature also shows a deviation from the calculated CI values. For example, comparing the published CIs for Norway from IEA (29.1\,gCO2/kWh \cite{IEA2021}), the top-down method (19.45\,gCO2/kWh), and the bottom-up method (16.31\,gCO2/kWh), all showing differences. This indicates that these differences are not the result of a single systematic bias, but rather are connected to multiple country-dependent causes and methodological differences (see Section~\ref{sec:discussion}).

\begin{table}[t]
\begin{tabular}{lrrr}
\toprule
Country & CI bottom-up & CI top-down &  Diff.  \\
& (gCO$_2$/kWh) &  (gCO$_2$/kWh) & (\%) \\
\midrule
SE & 15.36 & 13.29 & 13.0 \\
NO & 16.32 & 19.45 & $-19.0$ \\
FR & 37.16 & 56.57 & $-52.0$ \\
LT & 129.60 & 60.45 & 53.0 \\
AT & 136.27 & 106.16 & 22.0 \\
FI & 155.46 & 115.63 & 26.0 \\
SK & 231.71 & 144.24 & 38.0 \\
LV & 260.32 & 148.24 & 43.0 \\
DK & 223.17 & 200.42 & 10.0 \\
BE & 194.79 & 218.56 & $-12.0$ \\
GB & 287.08 & 256.56 & 11.0 \\
SI & 342.20 & 260.85 & 24.0 \\
IT & 323.79 & 262.45 & 19.0 \\
HU & 318.21 & 265.61 & 17.0 \\
ES & 342.77 & 291.10 & 15.0 \\
RO & 321.03 & 308.48 & 4.0 \\
PT & 317.05 & 326.01 & $-3.0$ \\
IE & 386.85 & 369.52 & 4.0 \\
DE & 435.15 & 423.38 & 3.0 \\
BG & 527.36 & 449.79 & 15.0 \\
NL & 469.08 & 465.48 & 1.0 \\
CZ & 479.78 & 470.43 & 2.0 \\
GR & 726.01 & 691.17 & 5.0 \\
CY & 879.69 & 703.98 & 20.0 \\
PL & 837.94 & 833.30 & 1.0 \\
EE & 895.14 & 943.07 & $-5.0$ \\ \bottomrule

\end{tabular}
\caption{Carbon intensity of electricity generation for EU countries for the year 2018. Bottom-up CIs are based on per unit emission and generation data, and per technology annual generation (see~Section~\ref{sec:bottomupmethod}), whereas top-down CIs are calculated using nationally reported energy balances and emissions (see~Section~\ref{sec:topdownmethod}).}
\label{tab:CI_countries}
\end{table}

\subsection{Dynamic carbon intensity signal}
\label{sec:CO2_signal}
\begin{figure*}[h]
	\centering
	\includegraphics[width=0.85\textwidth]{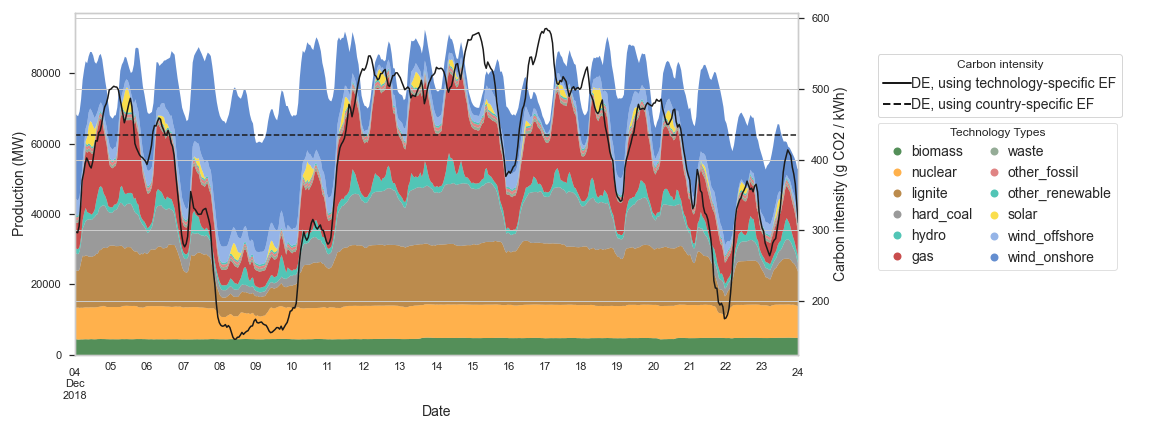}
	\caption{German generation mix in hourly resolution from December 4 until December 24, 2018. The dynamic hourly  carbon intensity of the generation mix is shown as a black line, compared to the static annually average carbon intensity as a dashed line.}
	\label{fig:CO2_signal_DE}
\end{figure*}

Publicly available datasets often contain only information about the average carbon intensity of electricity generation in a given country. Such a static value neglects the temporal variability of the generation mix. In contrast, per technology EFs allow to take into account this generation mix, leading to a dynamic assessment of the carbon intensity of power generation at a certain point in time. To illustrate this difference, the hourly generation mix and the associated carbon intensity in Germany and Poland for the week from December 4 to December 24, 2018 is depicted in Figure~\ref{fig:CO2_signal_DE}. The variability of the generation mix is reflected by the hourly carbon intensity of power generation $CI_t$ based on per technology EFs,
\begin{equation}
    CI_t=\frac{
    \sum_{\mathrm{tech}}G^{\mathrm{type}}_{t}(\mathrm{tech})\cdot EF(\mathrm{tech})}
    {
    \sum_{\mathrm{tech}}G^{\mathrm{type}}_{t}(\mathrm{tech})}~.
\end{equation}
Here, we have assigned zero CO$_2$ emissions to all non-fossil generation technologies. Figure~\ref{fig:CO2_signal_DE} indicates that the resulting hourly emission intensity signal $CI_t$ often significantly differs from the average per country carbon intensity $CI$.

\begin{figure*}[h]
	\centering
	\includegraphics[width=0.85\textwidth]{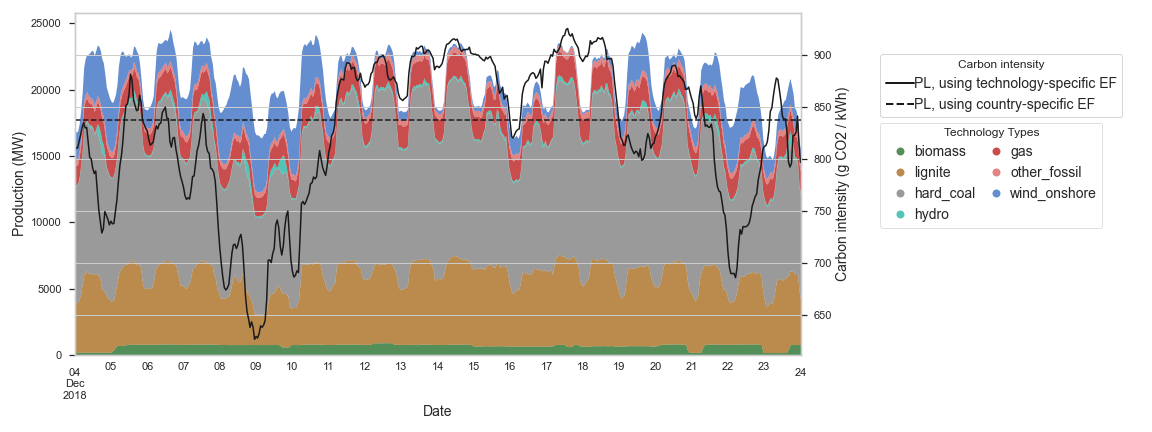}
	\caption{Poland generation mix in hourly resolution from December 4 until  December 24, 2018. The dynamic hourly  carbon intensity of the generation mix is shown as a black line, compared to the static annual average carbon intensity as a dashed line.}
	\label{fig:CO2_signal_PL}
\end{figure*}

\subsection{Carbon intensity duration curve}
\label{sec:CO2_durationcurve}
\begin{figure*}[h!]
	\centering
	\includegraphics[width=0.85\textwidth]{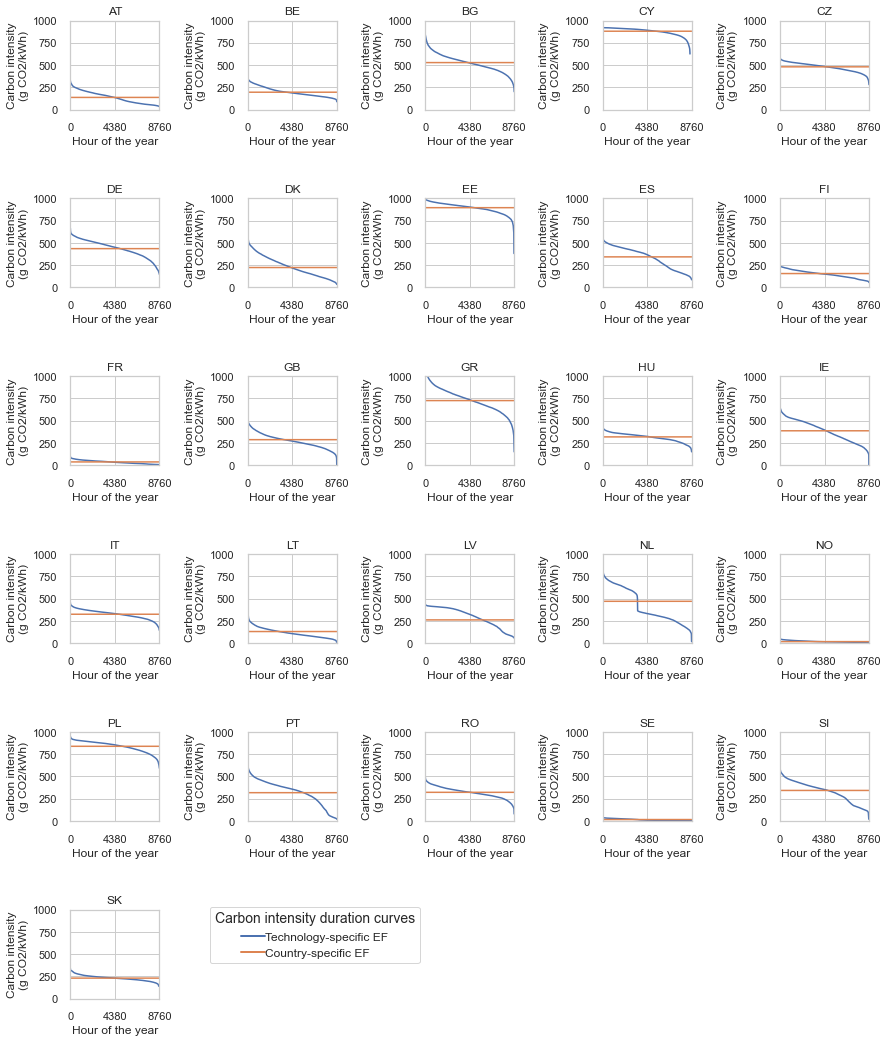}
	\caption{Carbon intensity duration curves for European countries for 2018. Hourly carbon intensity of electricity generation was calculated using the bottom-up method (see~Section\ref{sec:bottomupmethod}).}
	\label{fig:CO2_duration_curve}
\end{figure*}

To evaluate the variability of the carbon intensity of power generation across the year, carbon intensity duration curves for 26 European countries are represented in Figure~\ref{fig:CO2_duration_curve}. These display the hourly carbon intensity signals in descending order, for each country. The variability differs considerably between the countries, depending on the composition and usage of the national power generation fleet. Sweden, Norway and France, for instance, show a flat and low carbon intensity variability caused by their high and constant use of nuclear and hydropower. In contrast, Greece has a high and strongly varying CO$_2$ emission signal due to its power generation mix of approximately one-third of lignite, fossil gas and renewable energy, respectively~\cite{ENTSO-E2018}. For Poland, the low share of low-carbon generation causes the CI to remain high despite some variability around the average value. For countries with a significant share of intermittent renewable generation, such as Germany and Denmark (share of generation from wind and solar is approximately $25\%$ for Germany, and $52\,\%$ for Denmark), the associated variability in the generation mix translates into a wide range of carbon intensity values.

\section{Discussion}
\label{sec:discussion}

\subsection{Comparison of top-down and bottom-up carbon intensity results}

Since each calculation of the carbon intensity of electricity generation depends on the underlying assumptions and definitions, datasets, and specific methodological choices, there is not one ``correct'' result to compare other approaches against. To assess the influence of different factors in the calculation process, we thus focus on the comparison between the CIs resulting from the top-down and bottom-up method presented in Sections~\ref{sec:bottomupmethod} and~\ref{sec:topdownmethod}. While the top-down method  only yields static average annual values, the bottom-up method combines per technology EFs based on per unit data with generation time series to derive dynamic hourly carbon intensities. Table~\ref{tab:CI_countries} shows that the resulting average yearly CIs differ, depending on the method and country. Since these differences vary in magnitude and sign, it stands to reason that multiple underlying causes are involved. 

A first origin of discrepancy could be erroneous per unit EFs in the bottom-up method for some individual power plants. We have excluded implausible EFs from clear outliers (see Figure~\ref{fig:CO2_technology}), but for a further analysis, plant-specific EFs based on technical properties and fuel types would have to be evaluated. Although this is beyond the scope of our work, the suggested bottom-up method would allow to integrate corresponding results from future studies. A further uncertainty is the representativeness of the matched power plants for the generation technology categories of a given country. Although Figure~\ref{fig:deviation_CO2_unit_tech} indicates a minor influence of using per technology CFs compared to per unit EFs, generation units not contained or unmatched in the underlying EUTL and \mbox{ENTSO-E} datasets could affect the calculations. In most cases, we expect these influences to be minor due to the already significant coverage of our dataset (see Tab.~\ref{tab:EF_per_country_tech}), but an extension of the matching dataset or estimating the emissions of units that are not represented could provide a more accurate assessment. 

Using the bottom-up method, the annual average CI of a country is calculated through multiplication of hourly per technology generation time series with per country and technology EFs. Although these time series take into account consolidated data from \mbox{ENTSO-E}, the aggregated values differ considerably from the nationally reported electricity generation, as published by Eurostat. The inconsistencies are related to different coverages and categorizations in both datasets, originating from unequal reporting structures (for a discussion focusing on Germany, see~\cite{Schumacher2015}). For Italy, for example, it has been observed that most ``other'' generation reported in the \mbox{ENTSO-E} generation time series represents fossil gas~\cite{emberdata}. Since the associated ``other\_fossil'' in the bottom-up method (see Table~\ref{tab:technologies}) has a higher EF than fossil gas, the resulting CI is higher than the one calculated through the top-down method. Resolving this and related issues would involve establishing a consolidated dataset of hourly per country and technology generation time series, based on a comprehensive assessment of nationally and internationally published data and the underlying reporting methodologies. Such an endeavor is beyond the scope of this study, but, as in the case for per unit EFs, future revised generation time series can easily be integrated into the published bottom-up method.

As discussed in Section~\ref{sec:topdownmethod}, the top-down method employs energy flows as published by Eurostat to estimate the share of total reported emissions associated with electricity generation~\cite{EEA2021}. This process explicitly incorporates auto producers, for whom the production of electricity is not their principal activity. These units in general are excluded in the per unit generation data from \mbox{ENTSO-E}, so that this sector is underrepresented in the EFs contained in the bottom-up method. France, for example, has a share of $26\,\%$ of electricity-related energy input associated with auto producers, which could be the cause for the higher CI calculated through the top-down method, compared to bottom-up. This discrepancy could be resolved by either implementing corrections in the bottom-up EFs, or by removing auto producers from the top-down method.  As for the consolidation of the generation data,  a detailed per country analysis is necessary to account for this influence.

\subsection{Relation to results from the literature}
The German think tank Agora Energiewende employs per technology EFs from the Federal Environment Agency of Germany to calculate a close to real-time CI signal for the German electricity mix~\cite{Hein2020Agorameter-Dokumentation}. This real-time CI signal calculation is comparable to our "dynamic carbon intensity signal" calculation in Section~\ref{sec:CO2_signal}, combining per technology EFs with generation time series. Both methods consider only direct emissions, and imports and exports are excluded. Our method, however, allows a more flexible application to other countries since the necessary datasets are available for all European countries. Also, the calculation of per technology EFs is integrated in our method, whereas~\cite{Hein2020Agorameter-Dokumentation} uses data from other sources~\cite{Umweltbundesamt2021Entwicklung2020}. It should be noted that for 2018, these literature values are similar to the ones presented in this study  ($1\,090, 820, 370$ gCO$_2$/kWh for lignite, hard coal, gas in~\cite{Hein2020Agorameter-Dokumentation} compared to $1\,126, 934, 355$ gCO$_2$/kWh in Table~\ref{tab:EF_per_country_tech}). 

The company Tomorrow publishes real-time signals for the carbon intensity of electricity generation and consumption for European countries through the ``electricityMap''~\cite{electricitymap}. While historical data and the provisioning of an API is a fee-based offer, which constitutes the business model of the company, the electricityMap itself is open source and integrates contributions from the community~\cite{electricitymapgithub}. The generation-based CI is based on time series for per country and technology generation from various sources merged with EFs from the literature. These factors take into account emissions from the whole life-cycle of the plant, different from the direct emissions used in this study. The main source for these factors, according to~\cite{electricitymapgithub}, is the IPCC Fifth Assessment Report from 2014~\cite{IPCC2014}. Compared with the factors derived in this study, the literature values have a wider scope than direct emissions only, but do not discriminate between different countries. This neglects considerable country-wise differences in the per technology EFs, as they were shown in Table~\ref{tab:EF_per_country_tech} (for instance, $1\,126$ gCO$_2$/kWh for lignite in Germany compared to $1\,429$ gCO$_2$/kWh for lignite in Greece).

In~\cite{Vuarnoz2018TemporalGrid}, the hourly greenhouse gas emissions of the national electricity supply mix in Switzerland are analyzed, considering imports and exports with neighboring countries. For their analysis, the authors use CIs based on LCA studies. For the neighboring countries, a uniform EF is considered, neglecting hourly variations of the CI. As we have shown in Figure~\ref{fig:CO2_duration_curve}, these variations can be considerable, and neglecting them leads to imprecise results. 

In~\cite{Turconi2013LifeLimitations}, the authors review 167 LCA studies on electricity generation and the resulting emissions. The authors find that often different LCA methods are used and/or system boundaries are not chosen equally, which challenges the comparability of the derived results. In contrast, our developed bottom-up method allows to compare emissions of electricity generation from many technologies and countries without changing system boundaries and methods between technologies and countries. This method thus represents a valuable system-wide approach, complementary to more detailed and specific LCA methods.

Following a similar approach to the bottom-up method presented in this study,~\cite{Braeuer2020ComparingGermany} derive a CI signal based on power plant emission and generation data. The analysis is limited to the German energy system, with no detailed discussion of possible inconsistencies of the underlying data or the consideration of heat generation.

\section{Summary and conclusions}
\label{sec:conclusion}
Transparent and comprehensible emission factors (EFs) address the increasing need for dynamic grid carbon intensity signals for low-carbon system operation and emission accounting~\cite{Hamels2021}. In this contribution, we introduce a bottom-up framework to calculate per country and per technology direct EFs for European countries, based on publicly available per unit generation and emission data. The resulting EFs are merged with generation data to calculate the hourly carbon intensity of electricity generation. A comparison with results from a top-down approach based on emission and generation data from national statistics shows the feasibility of this approach, but also indicates the necessity of further consolidation of the underlying input data. In the proposed method, these extensions can be included on different levels, ranging from individual power plants to national generation time series or correction factors, taking into account not represented generation or emissions. The use of publicly available data and the publication of all code and auxiliary data, as well as the modularity of the approach facilitates building on the presented work in a flexible way.

\section*{Data availability}

The complete data processing and EF model is implemented in Python using the developing environment Jupyter Notebook \cite{Kluyver2016jupyter}. This facilitates transparent and accessible publication of the code, allowing other users to extend and adapt the method and integrate further data sources. The analysis and processing of the EFs was also done in Jupyter Notebook. All code is available on GitHub \cite{co2emissionsfactorsgithub}, with the processed input dataset \cite{unnewehr_jan_frederick_2021_5336486} and output dataset \cite{unnewehr_jan_frederick_2021_5603077} published on Zenodo .

\section*{Acknowledgements}

We thank Dave Jones and the team from Ember for supporting the work. Support with data processing and visualization from Paul Reggentin is acknowledged. We also thank Dr. Kevin Hausmann and Hauke Hermann for fruitful discussions.

\bibliographystyle{elsarticle-num}
\biboptions{sort}
\bibliography{literature_review_1}

\begin{thebibliography}{10}
\expandafter\ifx\csname url\endcsname\relax
  \def\url#1{\texttt{#1}}\fi
\expandafter\ifx\csname urlprefix\endcsname\relax\def\urlprefix{URL }\fi
\expandafter\ifx\csname href\endcsname\relax
  \def\href#1#2{#2} \def\path#1{#1}\fi

\bibitem{EU2019}
{European Commission}, {The European Green Deal COM(2019) 640} (2019).

\bibitem{EU2020}
{European Commission}, {Stepping up Europe’s 2030 climate ambition. Investing
  in a climate-neutral future for the benefit of our people OM(2020) 562}
  (2020).

\bibitem{Hamels2021}
S.~Hamels, E.~Himpe, J.~Laverge, M.~Delghust, K.~Van~den Brande, A.~Janssens,
  J.~Albrecht, {The use of primary energy factors and CO2 intensities for
  electricity in the European context - A systematic methodological review and
  critical evaluation of the contemporary literature}, Renewable and
  Sustainable Energy Reviews 146 (2021) 111182.
\newblock \href {https://doi.org/10.1016/j.rser.2021.111182}
  {\path{doi:10.1016/j.rser.2021.111182}}.

\bibitem{Clauss2019}
J.~Clau{\ss}, S.~Stinner, C.~Solli, K.~B. Lindberg, H.~Madsen, L.~Georges,
  {Evaluation Method for the Hourly Average CO$_{2\text{eq.}}$ Intensity of the
  Electricity Mix and Its Application to the Demand Response of Residential
  Heating}, Energies 12 (2019) 1345.
\newblock \href {https://doi.org/10.3390/en12071345}
  {\path{doi:10.3390/en12071345}}.

\bibitem{Alavijeh2020}
N.~M. Alavijeh, D.~Steen, Z.~Norwood, L.~A. Tuan, C.~Agathokleous,
  {Cost-Effectiveness of Carbon Emission Abatement Strategies for a Local
  Multi-Energy System-A Case Study of Chalmers University of Technology
  Campus}, Energies 13 (2020) 1626.
\newblock \href {https://doi.org/10.3390/en13071626}
  {\path{doi:10.3390/en13071626}}.

\bibitem{Google2021}
R.~Koningstein,
  \href{https://www.blog.google/outreach-initiatives/sustainability/carbon-aware-computing-location/}{{We
  now do more computing where there’s cleaner energy}} (2021).
\newline\urlprefix\url{https://www.blog.google/outreach-initiatives/sustainability/carbon-aware-computing-location/}

\bibitem{RMI2021}
M.~Dyson, S.~Shah, C.~Teplic,
  \href{http://www.rmi.org/insight/clean-power-by-the-hour}{{Clean Power by the
  Hour: Assessing the Costs and Emissions Impacts of Hourly Carbon-Free Energy
  Procurement Strategies}}, Tech. rep., RMI (2021).
\newline\urlprefix\url{http://www.rmi.org/insight/clean-power-by-the-hour}

\bibitem{climatetrace}
\href{https://www.climatetrace.org/}{{Climate Trace}}, accessed: 2021-08-02
  (2020).
\newline\urlprefix\url{https://www.climatetrace.org/}

\bibitem{EEA2021}
{European Environment Agency (EEA)},
  \href{https://www.eea.europa.eu/data-and-maps/indicators/overview-of-the-electricity-production-3/assessment-1}{{Indicator
  Assessment: Greenhouse gas emission intensity of electricity generation in
  Europe}} (2021).
\newline\urlprefix\url{https://www.eea.europa.eu/data-and-maps/indicators/overview-of-the-electricity-production-3/assessment-1}

\bibitem{1992UNITEDNATIONS}
\href{https://unfccc.int/resource/docs/convkp/conveng.pdf}{{United Nations
  Framework Convention on Climate Change}} (1992).
\newline\urlprefix\url{https://unfccc.int/resource/docs/convkp/conveng.pdf}

\bibitem{Eggleston20062006Inventories}
S.~Eggleston, L.~Buendia, K.~Miwa, T.~Ngara, K.~Tanabe,
  \href{https://www.ipcc-nggip.iges.or.jp/public/2006gl/}{{2006 IPCC guidelines
  for national greenhouse gas inventories}} (2006).
\newline\urlprefix\url{https://www.ipcc-nggip.iges.or.jp/public/2006gl/}

\bibitem{Turconi2013LifeLimitations}
R.~Turconi, A.~Boldrin, T.~Astrup, {Life cycle assessment (LCA) of electricity
  generation technologies: Overview, comparability and limitations}, Renewable
  and Sustainable Energy Reviews 28 (2013) 555--565.
\newblock \href {https://doi.org/10.1016/j.rser.2013.08.013}
  {\path{doi:10.1016/j.rser.2013.08.013}}.

\bibitem{Astudillo2017LifeOpportunities}
M.~F. Astudillo, K.~Treyer, C.~Bauer, P.~O. Pineau, M.~B. Amor, {Life cycle
  inventories of electricity supply through the lens of data quality: exploring
  challenges and opportunities}, The International Journal of Life Cycle
  Assessment 22~(3) (2017) 374--386.
\newblock \href {https://doi.org/10.1007/s11367-016-1163-0}
  {\path{doi:10.1007/s11367-016-1163-0}}.

\bibitem{Itten2012LifeGrid}
R.~Itten, R.~Frischknecht, M.~Stucki, P.~Scherrer, I.~Psi,
  \href{http://esu-services.ch/fileadmin/download/publicLCI/itten-2012-electricity-mix.pdf}{{Life
  cycle inventories of electricity mixes and grid}} (2012).
\newline\urlprefix\url{http://esu-services.ch/fileadmin/download/publicLCI/itten-2012-electricity-mix.pdf}

\bibitem{Tranberg2019Real-timeMarkets}
B.~Tranberg, O.~Corradi, B.~Lajoie, T.~Gibon, I.~Staffell, G.~B. Andresen,
  {Real-time carbon accounting method for the European electricity markets},
  Energy Strategy Reviews 26 (12 2019).
\newblock \href {https://doi.org/10.1016/j.esr.2019.100367}
  {\path{doi:10.1016/j.esr.2019.100367}}.

\bibitem{Spork2015IncreasingFactors}
C.~C. Spork, A.~Chavez, X.~Gabarrell~Durany, M.~K. Patel,
  G.~Villalba~M{\'{e}}ndez, {Increasing Precision in Greenhouse Gas Accounting
  Using Real-Time Emission Factors}, Journal of Industrial Ecology 19~(3)
  (2015) 380--390.
\newblock \href {https://doi.org/10.1111/jiec.12193}
  {\path{doi:10.1111/jiec.12193}}.

\bibitem{Noussan2018PerformanceItaly}
M.~Noussan, R.~Roberto, B.~Nastasi, {Performance Indicators of Electricity
  Generation at Country Level -- The Case of Italy}, Energies (2018).
\newblock \href {https://doi.org/10.3390/en11030650}
  {\path{doi:10.3390/en11030650}}.

\bibitem{Kopsakangas-Savolainen2017Hourly-basedEmissions}
M.~Kopsakangas-Savolainen, M.~K. Mattinen, K.~Manninen, A.~Nissinen,
  {Hourly-based greenhouse gas emissions of electricity -– cases
  demonstrating possibilities for households and companies to decrease their
  emissions}, Journal of Cleaner Production 153 (2017) 384--396.
\newblock \href {https://doi.org/10.1016/j.jclepro.2015.11.027}
  {\path{doi:10.1016/j.jclepro.2015.11.027}}.

\bibitem{Vuarnoz2018TemporalGrid}
D.~Vuarnoz, T.~Jusselme, {Temporal variations in the primary energy use and
  greenhouse gas emissions of electricity provided by the Swiss grid}, Energy
  161 (2018) 573--582.
\newblock \href {https://doi.org/10.1016/j.energy.2018.07.087}
  {\path{doi:10.1016/j.energy.2018.07.087}}.

\bibitem{Marrasso2019ElectricBasis}
E.~Marrasso, C.~Roselli, M.~Sasso, {Electric efficiency indicators and carbon
  dioxide emission factors for power generation by fossil and renewable energy
  sources on hourly basis}, Energy Conversion and Management 196 (2019)
  1369--1384.
\newblock \href {https://doi.org/10.1016/j.enconman.2019.06.079}
  {\path{doi:10.1016/j.enconman.2019.06.079}}.

\bibitem{Braeuer2020ComparingGermany}
F.~Braeuer, R.~Finck, R.~McKenna, {Comparing empirical and model-based
  approaches for calculating dynamic grid emission factors: An application to
  CO2-minimizing storage dispatch in Germany}, Journal of Cleaner Production
  266 (2020) 121588.
\newblock \href {https://doi.org/10.1016/j.jclepro.2020.121588}
  {\path{doi:10.1016/j.jclepro.2020.121588}}.

\bibitem{Staffell2017MeasuringElectricity}
I.~Staffell, {Measuring the progress and impacts of decarbonising British
  electricity}, Energy Policy 102 (2017) 463--475.
\newblock \href {https://doi.org/10.1016/j.enpol.2016.12.037}
  {\path{doi:10.1016/j.enpol.2016.12.037}}.

\bibitem{Hein2020Agorameter-Dokumentation}
F.~Hein, H.~Hermann, \href{www.agora-energiewende.de}{{Agorameter --
  Dokumentation Version 10}} (2020).
\newline\urlprefix\url{www.agora-energiewende.de}

\bibitem{Dixit2014CalculatingSectors}
M.~K. Dixit, C.~H. Culp, J.~L. Fernandez-Solis, {Calculating primary energy and
  carbon emission factors for the United States energy sectors}, RSC Advances
  4~(97) (2014) 54200--54216.
\newblock \href {https://doi.org/10.1039/c4ra08989h}
  {\path{doi:10.1039/c4ra08989h}}.

\bibitem{Moro2018ElectricityVehicles}
A.~Moro, L.~Lonza, {Electricity carbon intensity in European Member States:
  Impacts on GHG emissions of electric vehicles}, Transportation Research Part
  D: Transport and Environment 64 (2018) 5--14.
\newblock \href {https://doi.org/10.1016/J.TRD.2017.07.012}
  {\path{doi:10.1016/J.TRD.2017.07.012}}.

\bibitem{Arciniegas2018TradeoffsOperation}
L.~M. Arciniegas, E.~Hittinger, {Tradeoffs between revenue and emissions in
  energy storage operation}, Energy 143 (2018) 1--11.
\newblock \href {https://doi.org/10.1016/j.energy.2017.10.123}
  {\path{doi:10.1016/j.energy.2017.10.123}}.

\bibitem{Ang2016CarbonAnalysis}
B.~W. Ang, B.~Su, {Carbon emission intensity in electricity production: A
  global analysis}, Energy Policy 94 (2016) 56--63.
\newblock \href {https://doi.org/10.1016/j.enpol.2016.03.038}
  {\path{doi:10.1016/j.enpol.2016.03.038}}.

\bibitem{Qu2017}
S.~Qu, H.~Wang, S.~Liang, A.~M. Shapiro, S.~Suh, S.~Sheldon, O.~Zik, H.~Fang,
  M.~Xu, {A Quasi-Input-Output model to improve the estimation of emission
  factors for purchased electricity from interconnected grids}, Applied Energy
  200 (2017) 249--259.
\newblock \href {https://doi.org/10.1016/J.APENERGY.2017.05.046}
  {\path{doi:10.1016/J.APENERGY.2017.05.046}}.

\bibitem{Zafirakis2015}
D.~Zafirakis, K.~J. Chalvatzis, G.~Baiocchi, {Embodied CO2 emissions and
  cross-border electricity trade in Europe: Rebalancing burden sharing with
  energy storage}, Applied Energy 143 (2015) 283--300.
\newblock \href {https://doi.org/10.1016/J.APENERGY.2014.12.054}
  {\path{doi:10.1016/J.APENERGY.2014.12.054}}.

\bibitem{Odeh2008LifePlants}
N.~A. Odeh, T.~T. Cockerill, {Life cycle analysis of UK coal fired power
  plants}, Energy Conversion and Management 49~(2) (2008) 212--220.
\newblock \href {https://doi.org/10.1016/j.enconman.2007.06.014}
  {\path{doi:10.1016/j.enconman.2007.06.014}}.

\bibitem{Umweltbundesamt2021Entwicklung2020}
P.~Icha, {Entwicklung der spezifischen Kohlendioxid-Emissionen des deutschen
  Strommix in den Jahren 1990 - 2019}, Tech. Rep. Climate Change 45/2021,
  {Umweltbundesamt} (2021).

\bibitem{USEIA}
\href{https://www.eia.gov/environment/emissions/co2_vol_mass.php}{{Carbon
  Dioxide Emissions Coefficients}}, accessed: 2021-08-03 (2016).
\newline\urlprefix\url{https://www.eia.gov/environment/emissions/co2_vol_mass.php}

\bibitem{electricitymap}
Tomorrow, {electricityMap}, \url{https://www.electricitymap.org/map}, accessed:
  2021-07-28.

\bibitem{electricitymapgithub}
{electricityMap GitHub repository},
  \url{https://github.com/tmrowco/electricitymap-contrib}, accessed:
  2021-07-28.

\bibitem{IPCC2014}
S.~Schlömer, T.~Bruckner, L.~Fulton, E.~Hertwich, A.~McKinnon, D.~Perczyk,
  J.~Roy, R.~Schaeffer, R.~Sims, P.~Smith, R.~Wiser, Annex iii:
  Technology-specific cost and performance parameter, in: Climate Change 2014:
  Mitigation of Climate Change. Contribution of Working Group III to the Fifth
  Assessment Report of the Intergovernmental Panel on Climate Change, 2014.

\bibitem{deChalendar2019}
J.~A. de~Chalendar, J.~Taggart, S.~M. Benson, {Tracking emissions in the US
  electricity system}, Proceedings of the National Academy of Sciences 116~(51)
  (2019) 25497--25502.
\newblock \href {https://doi.org/10.1073/pnas.1912950116}
  {\path{doi:10.1073/pnas.1912950116}}.

\bibitem{Schaefer2020}
M.~Schäfer, B.~Tranberg, D.~Jones, A.~Weidlich, Tracing carbon dioxide
  emissions in the european electricity markets, in: 17th International
  Conference on the European Energy Market (EEM), 2020, pp. 1--6.
\newblock \href {https://doi.org/10.1109/EEM49802.2020.9221928}
  {\path{doi:10.1109/EEM49802.2020.9221928}}.

\bibitem{2020ENTSO-EUnit}
\href{https://transparency.entsoe.eu/generation/r2/actualGenerationPerGenerationUnit/show}{{ENTSO-E
  Transparency Platform, Actual Generation per Generation Unit}}, accessed:
  2021-07-28.
\newline\urlprefix\url{https://transparency.entsoe.eu/generation/r2/actualGenerationPerGenerationUnit/show}

\bibitem{2020ENTSO-EType}
\href{https://transparency.entsoe.eu/generation/r2/actualGenerationPerProductionType/show}{{ENTSO-E
  Transparency Platform, Actual Generation per Production Type}}, accessed:
  2021-07-28.
\newline\urlprefix\url{https://transparency.entsoe.eu/generation/r2/actualGenerationPerProductionType/show}

\bibitem{EnergyEurostat}
Eurostat,
  \href{https://ec.europa.eu/eurostat/web/energy/data/energy-balances}{{Energy
  balances}}, accessed: 2021-07-28.
\newline\urlprefix\url{https://ec.europa.eu/eurostat/web/energy/data/energy-balances}

\bibitem{EUROPALog}
{European Commission}, \href{https://ec.europa.eu/clima/ets/}{{European Union
  Transaction Log}}, accessed: 2021-07-28.
\newline\urlprefix\url{https://ec.europa.eu/clima/ets/}

\bibitem{NationalAgency}
{European Environment Agency (EEA)},
  \href{https://www.eea.europa.eu/data-and-maps/data/national-emissions-reported-to-the-unfccc-and-to-the-eu-greenhouse-gas-monitoring-mechanism-17}{{National
  emissions reported to the UNFCCC and to the EU Greenhouse Gas Monitoring
  Mechanism}} (2021).
\newline\urlprefix\url{https://www.eea.europa.eu/data-and-maps/data/national-emissions-reported-to-the-unfccc-and-to-the-eu-greenhouse-gas-monitoring-mechanism-17}

\bibitem{ENTSO-E2018}
{ENTSO-E}, {Statistical Factsheet 2018}, Tech. rep. (2019).

\bibitem{Hirth2018}
L.~Hirth, J.~M{\"{u}}hlenpfordt, M.~Bulkeley, {The ENTSO-E Transparency
  Platform -– A review of Europe's most ambitious electricity data platform},
  Applied Energy 225~(April) (2018) 1054--1067.
\newblock \href {https://doi.org/10.1016/j.apenergy.2018.04.048}
  {\path{doi:10.1016/j.apenergy.2018.04.048}}.

\bibitem{AGEnergiebilanzene.V.2020Stromerzeugung2020}
{AG Energiebilanzen e.V.}, \href{https://ag-energiebilanzen.de}{{Stromerzeugung
  nach Energietr{\"{a}}gern 1990 - 2020}}, Tech. rep. (2021).
\newline\urlprefix\url{https://ag-energiebilanzen.de}

\bibitem{co2emissionsfactorsgithub}
{CO2 emissions factors GitHub repository},
  \url{https://github.com/INATECH-CIG/CO2_emissions_factors}, accessed:
  2021-08-12.

\bibitem{EUAction}
{European Commission}, \href{https://ec.europa.eu/clima/policies/ets_en}{{EU
  Emissions Trading System (EU ETS)}}, accessed: 2021-07-28.
\newline\urlprefix\url{https://ec.europa.eu/clima/policies/ets_en}

\bibitem{EuropeanCommission2015EUHandbook}
{European Commission},
  \href{https://ec.europa.eu/clima/sites/clima/files/docs/ets_handbook_en.pdf}{{EU
  ETS Handbook}}, Tech. rep. (2015).
\newline\urlprefix\url{https://ec.europa.eu/clima/sites/clima/files/docs/ets_handbook_en.pdf}

\bibitem{gemwiki}
\href{https://www.gem.wiki}{{Global Energy Monitor Wiki}}, accessed:
  2021-08-03.
\newline\urlprefix\url{https://www.gem.wiki}

\bibitem{JRCPPDB}
K.~Kanellopoulos, M.~{De Felice}, I.~Hidalgo, A.~Bocin, A.~Uihlein, {The Joint
  Research Centre Power Plant Database (JRC-PPDB) Version 0.9}, Tech. rep.,
  Joint Research Centre (2019).

\bibitem{Gotzens2019}
F.~Gotzens, H.~Heinrichs, J.~Hörsch, F.~Hofmann, Performing energy modelling
  exercises in a transparent way - the issue of data quality in power plant
  databases, Energy Strategy Reviews 23 (2019) 1--12.
\newblock \href {https://doi.org/https://doi.org/10.1016/j.esr.2018.11.004}
  {\path{doi:https://doi.org/10.1016/j.esr.2018.11.004}}.

\bibitem{OPSD2020}
\href{https://doi.org/10.25832/conventional_power_plants/2020-10-01}{{Data
  Package Conventional power plants. Version 2020-10-01.}}, accessed:
  2021-08-03 (2020).
\newline\urlprefix\url{https://doi.org/10.25832/conventional_power_plants/2020-10-01}

\bibitem{unnewehr_jan_frederick_2021_5603077}
J.~F. Unnewehr, {Output data for Open-data based carbon emission intensity
  signals for electricity generation in European countries -- top down vs.
  bottom up approach} (Oct. 2021).
\newblock \href {https://doi.org/10.5281/zenodo.5603077}
  {\path{doi:10.5281/zenodo.5603077}}.

\bibitem{NREL2021}
NREL, \href{https://www.nrel.gov/docs/fy21osti/80580.pdf}{{Life Cycle
  Greenhouse Gas Emissions from Electricity Generation: Update}} (2021).
\newline\urlprefix\url{https://www.nrel.gov/docs/fy21osti/80580.pdf}

\bibitem{Li2009NOxStaging}
S.~Li, T.~Xu, S.~Hui, X.~Wei, {NOx emission and thermal efficiency of a 300 MWe
  utility boiler retrofitted by air staging}, Applied Energy 86~(9) (2009)
  1797--1803.
\newblock \href {https://doi.org/10.1016/j.apenergy.2008.12.032}
  {\path{doi:10.1016/j.apenergy.2008.12.032}}.

\bibitem{IEA2021}
IEA,
  \href{https://www.iea.org/data-and-statistics/data-product/emissions-factors-2021}{{Emissions
  Factors 2021}} (2021).
\newline\urlprefix\url{https://www.iea.org/data-and-statistics/data-product/emissions-factors-2021}

\bibitem{Schumacher2015}
M.~Schumacher, L.~Hirth, {How much Electricity do we Consume?: A Guide to
  German and European Electricity Consumption and Generation Data}, Working
  paper, Fondazione Eni Enrico Mattei (FEEM) (2015).

\bibitem{emberdata}
{Ember}, \href{https://ember-climate.org/european-electricity-transition/}{{EU
  Electricity Data June 2021}}, accessed: 2021-08-02 (2021).
\newline\urlprefix\url{https://ember-climate.org/european-electricity-transition/}

\bibitem{Kluyver2016jupyter}
T.~Kluyver, B.~Ragan-Kelley, F.~P{\'{e}}rez, B.~Granger, M.~Bussonnier,
  J.~Frederic, K.~Kelley, J.~Hamrick, J.~Grout, S.~Corlay, P.~Ivanov, D.~Avila,
  S.~Abdalla, C.~Willing, {Jupyter Notebooks -- a publishing format for
  reproducible computational workflows}, in: F.~Loizides, B.~Schmidt (Eds.),
  Positioning and Power in Academic Publishing: Players, Agents and Agendas,
  IOS Press, 2016, pp. 87--90.

\bibitem{unnewehr_jan_frederick_2021_5336486}
J.~F. Unnewehr, {Input data for Open-data based carbon emission intensity
  signals for electricity generation in European countries -- top down vs.
  bottom up approach} (Aug. 2021).
\newblock \href {https://doi.org/10.5281/zenodo.5336486}
  {\path{doi:10.5281/zenodo.5336486}}.

\end{thebibliography}

\end{document}